\documentclass[letterpaper,english,prd, reprint, longbibliography, english, sort&compress, showpacs, showkeys]{revtex4-1}
\pdfoutput=1
\usepackage[T1]{fontenc}
\usepackage[latin9]{inputenc}
\usepackage{babel}
\usepackage{amsmath}
\usepackage{amssymb}
\usepackage{graphicx}
\usepackage[unicode=true,pdfusetitle,
 bookmarks=true,bookmarksnumbered=false,bookmarksopen=false,
 breaklinks=false,pdfborder={0 0 1},backref=false,colorlinks=false]
 {hyperref}

\makeatletter

\pdfpageheight\paperheight
\pdfpagewidth\paperwidth

\newcommand{\noun}[1]{\textsc{#1}}

\@ifundefined{showcaptionsetup}{}{%
 \PassOptionsToPackage{caption=false}{subfig}}
\usepackage{subfig}
\makeatother

\begin{document}

\title{Soft symmetry improvement of two particle irreducible effective actions}

\author{Michael J. Brown}
\email{michael.brown6@my.jcu.edu.au}

\author{Ian B. Whittingham}

\affiliation{College of Science and Engineering, James Cook University, Townsville
4811, Australia}

\date{\today}
\begin{abstract}
Two particle irreducible effective actions (2PIEAs) are valuable non-perturbative
techniques in quantum field theory; however, finite truncations of
them violate the Ward identities (WIs) of theories with spontaneously
broken symmetries. The symmetry improvement (SI) method of Pilaftsis
and Teresi attempts to overcome this by imposing the WIs as constraints
on the solution; however the method suffers from the non-existence
of solutions in linear response theory and in certain truncations
in equilibrium. Motivated by this, we introduce a new method called
\emph{soft symmetry improvement} (SSI) which relaxes the constraint.
Violations of WIs are allowed but punished in a least-squares implementation
of the symmetry improvement idea. A new parameter $\xi$ controls
the strength of the constraint. The method interpolates between the
unimproved ($\xi\to\infty$) and SI ($\xi\to0$) cases and the hope
is that practically useful solutions can be found for finite $\xi$.
We study the SSI-2PIEA for a scalar $\mathrm{O}\left(N\right)$ model
in the Hartree-Fock approximation. We find that the method is IR sensitive:
the system must be formulated in finite volume $\mathrm{V}$ and temperature
$T=\beta^{-1}$ and the $\mathrm{V}\beta\to\infty$ limit taken carefully.
Three distinct limits exist. Two are equivalent to the unimproved
2PIEA and SI-2PIEA respectively, and the third is a new limit where
the WI is satisfied but the phase transition is strongly first order
and solutions can fail to exist depending on $\xi$. Further, these
limits are disconnected from each other; there is no smooth way to
interpolate from one to another. These results suggest that any potential
advantages of SSI methods, and indeed any application of (S)SI methods
out of equilibrium, must occur in finite volume.
\end{abstract}


\maketitle

\section{\label{sec:Introduction}Introduction}

There is a growing interest in techniques for non-perturbative and
non-equilibrium quantum field theories. Potential applications for
new methods range from cold atoms to cosmology (see e.g. \citep{Berges2015}).
Recent progress on topics such as the dynamics of non-equilibrium
critical points and phase transitions has come from the development
of $n$-particle irreducible effective action ($n$PIEA; $n=1,2,3,\ldots$)
methods. These methods have a long history. The 1PIEA was introduced
by Goldstone, Salam and Weinberg \citep{Goldstone1962} and \citet{Jona-Lasinio1964}.
The 2PIEA was introduced independently by several authors \citep{Lee1960,Luttinger1960,Baym1962}
and finally received its modern formulation by Cornwall, Jackiw and
Tomboulis \citep{Cornwall1974}. This method has seen widespread use
in both condensed matter and fundamental physics (see, e.g. \citep{Berges2004,Berges2015}
for fairly recent reviews). \citet{DeDominicis1964} then realized
that these were special cases of a general formalism for arbitrary
$n$. This work was then extended by others \citep{Vasiliev1998,Kleinert1982,Haymaker1991,Berges2004},
but the practical use of effective actions for $n\geq3$ remains minimal,
largely due to difficulties with the renormalization of physically
interesting theories.

$n$PIEAs can be thought of as generalizations of mean field theory
which are (a) elegant, (b) general, (c) in principle exact, and (d)
have been promoted for their applicability to non-equilibrium situations
(see, e.g., \citep{Berges2015} and references therein for extensive
discussion of all these points). Non-perturbative methods are essential
in non-equilibrium QFT because secular terms (i.e. terms which grow
without bound over time) in the time evolution equations invalidate
perturbation theory. $n$PIEAs with $n>1$ achieve the required non-perturbative
resummation in a manifestly self-consistent way which can be derived
from first principles. ``In principle exact'' here means that the
$n$PIEA equations of motion are exactly equivalent to the original
non-perturbative definition of the quantum field theory. The only
necessary approximation is in the numerical solution of these equations.
The resulting equations of motion are also useful in equilibrium because
many-body effects are included self-consistently. ``General'' means
that the methods are applicable in principle to any quantum field
theory whatsoever (although in a theory with many fields or with large
$n$ the resulting $n$PIEA could be very bulky). Finally, ``elegant''
here means that few conceptually new elements are needed in the formulation
of $n$PIEAs in addition to the usual terms of textbook quantum field
theory. The complication is mainly of a technical, not conceptual,
nature. To our knowledge no other techniques satisfy all of these
criteria.

$n$PIEA methods work by recasting perturbation theory as a variational
method. Instead of working with standard Feynman diagrams built from
bare propagators and vertices, one works with a reduced set of Feynman
diagrams built from the \emph{exact} mean field $\varphi$, propagators
$\Delta$ and vertex functions $V^{\left(3\right)}$, $V^{\left(4\right)},\ldots,V^{\left(n\right)}$.
These quantities are determined by solving equations of motion $\delta\Gamma^{\left(n\right)}/\delta\varphi=\delta\Gamma^{\left(n\right)}/\delta\Delta=0$
etc. The $\Gamma^{\left(n\right)}$ functionals are themselves built
from $\varphi$, $\Delta$, $V^{\left(3\right)}$ and so on. The $\Gamma^{\left(n\right)}$
and accompanying equations of motion are exactly equivalent to the
original quantum field theory, but are sensitive to physical effects
which are invisible to perturbation theory. Furthermore, this ability
to capture non-perturbative physics is competitive with or exceeds
other standard resummation methods such as Borel-Pad\'{e} summation,
at least in a toy model where exact solutions are available as a benchmark
\citep{Brown2015}.

An unfortunate practical difficulty faced by would-be users of $n$PIEAs
is that, once truncated to finite order, solutions of the equations
of motion derived from $\Gamma^{\left(n>1\right)}$ no longer obey
the expected symmetry properties (i.e. Ward identities or WIs) which
are obeyed by the exact solution, even if the truncation is manifestly
invariant. This occurs simply because there is no guarantee that the
pattern of partial resummations encoded in an approximation to $\Gamma^{\left(n\right)}$
will respect the order by order cancellations required to fully maintain
the WIs. The most obvious effect of this is that Goldstone bosons
are unphysically massive and the symmetry breaking phase transition
is incorrectly predicted in models of spontaneous symmetry breaking
treated within the Hartree-Fock approximation (see, e.g. \citep{Pilaftsis2013}
and references therein). The use of higher order truncations can cure
this problem but more subtle symmetry violating effects still occur.
Similar remarks apply for gauge theories, where an unphysical gauge
dependence remains in quantities that should be physical.

Several methods have been advocated in the literature to combat this
problem, though none are without flaws. For example, the widely used
external propagator method \citep{VanHees2002} is not fully self-consistent:
\emph{after} the variational solution is found, ``external'' correlation
functions are constructed which \emph{do} satisfy the the WIs. However,
the incorrect variational solutions are still the ones used in the
self-consistent step. As a result, more subtle problems such as violations
of unitarity persist. \citet{Ivanov2005} developed a gapless version
of the 2PIEA in the Hartree-Fock approximation which restores the
second order phase transition and Goldstone theorem, but requires
the addition of an ad hoc correction term. There is not, as far as
we know, any first principles motivation for the scheme or any systematic
way of extending it. \citet{Leupold2007} discusses the use of nonlinear
representations, which restores the symmetry at the expense of requiring
nonpolynomial Lagrangians. \citet{Pilaftsis2013} introduced a promising
method called symmetry improvement (SI), which imposes the WIs directly
as constraints on the solution through Lagrange multipliers. SI has
been applied with some success with the SI-2PIEA \citep{Pilaftsis2013,Mao2014,Lu2015,Pilaftsis2015,Pilaftsis2015a}
and extended to the SI-3PIEA \citep{Brown2015a}, however the method
is inconsistent out of equilibrium (at least at the linear response
level) \citep{Brown_2016} and sometimes solutions fail to exist due
to the constraint causing a renormalization group defying coupling
between short and long distance physics \citep{Marko2016}. Considering
that the symptom in both cases is the non-existence of solutions,
and that the constraint in the SI method is singular and requires
some careful treatment to begin with, it is reasonable to suspect
that the culprit may be that the method is over-constraining. This
motivates the investigation of whether it is possible to generalize
the SI method and at the same time allow the solutions more freedom.
That is what this paper does.

We introduce a new method which we call \emph{soft symmetry improvement}
(SSI) which relaxes the constraint. Violations of WIs are allowed
but punished in the solution of the SSI-$n$PIEA. The method is essentially
a least-squares implementation of the symmetry improvement idea. A
new parameter, the stiffness $\xi$, controls the strength of the
constraint. The method interpolates between the unimproved ($\xi\to\infty$)
and SI ($\xi\to0$) cases and the hope is that practically useful
solutions can be found for finite $\xi$. We study the SSI-2PIEA for
a scalar $\mathrm{O}\left(N\right)$ model in the Hartree-Fock approximation.
We find that the method is IR sensitive: the system must be formulated
in finite volume $\mathrm{V}$ and temperature $T=\beta^{-1}$ and
the $\mathrm{V}\beta\to\infty$ limit must be taken carefully. Three
distinct limits exist. Two are equivalent to the unimproved 2PIEA
and SI-2PIEA respectively, and the third is a new limit where the
WI is satisfied but the phase transition is strongly first order and
solutions can fail to exist depending on $\xi$. Further, these limits
are disconnected from each other; there is no smooth way to interpolate
from one to another. These results suggest that any potential advantages
of SSI methods (and any consideration of (S)SI out of equilibrium)
\emph{must occur in finite volume}.

The structure of this paper is as follows. Following this introduction,
section \ref{sec:Soft-Symmetry-Improvement} introduces the SSI formalism.
Then, in section \ref{sec:Renormalisation-of-HF} the SSI-2PIEA is
renormalized in the Hartree-Fock approximation at finite $\mathrm{V}\beta$.
Solutions are then found in section \ref{sec:Solution-in-the-infinite-volume-limit}
with careful consideration of the various $\mathrm{V}\beta\to\infty$
limits. Finally, we discuss our results in section \ref{sec:Discussion}.
The notation agrees with our previous papers \citep{Brown2015a,Brown_2016}
except where noted. In particular, the deWitt summation convention
is used, i.e. sums over repeated indices imply integrations over corresponding
spacetime arguments.

\section{\label{sec:Soft-Symmetry-Improvement}Soft Symmetry Improvement of
2PIEA}

The soft symmetry improved 2PIEA is a modification of the 2PIEA defined
for theories with an internal symmetry. In order to have a concrete
example we use the $\mathrm{O}\left(N\right)$ symmetric scalar $\left(\phi^{2}\right)^{2}$
theory discussed in our previous papers \citep{Brown2015a,Brown_2016}.
We will focus on the spontaneous symmetry breaking regime where the
field has a non-zero expectation value $\varphi_{a}=\left\langle \phi_{a}\right\rangle =\left(0,\ldots,0,v\right)$,
a ``Higgs'' boson with mass $m_{H}$ and $N-1$ massless Goldstone
bosons. The definition of the SSI-2PIEA can be motivated by starting
with the standard 2PIEA $\Gamma\left[\varphi,\Delta\right]$ (suppressing
indices and spacetime arguments where these just clutter) and the
trivial identity
\begin{equation}
\exp\left(\frac{i}{\hbar}\Gamma\left[\varphi,\Delta\right]\right)=\int\mathcal{D}\phi\ \delta\left(\phi-\varphi\right)\exp\left(\frac{i}{\hbar}\Gamma\left[\phi,\Delta\right]\right).
\end{equation}

The usual symmetry improved action $\Gamma^{\mathrm{SI}}\left[\varphi,\Delta\right]$
is then obtained by inserting a delta function
\begin{align}
\exp\left(\frac{i}{\hbar}\Gamma^{\mathrm{SI}}\left[\varphi,\Delta\right]\right) & =N\int\mathcal{D}\phi\ \delta\left(\phi-\varphi\right)\nonumber \\
 & \times\exp\left(\frac{i}{\hbar}\Gamma\left[\phi,\Delta\right]\right)\delta\left(\mathcal{W}\left[\phi,\Delta\right]\right),
\end{align}
where the Ward identity is \citep{Brown2015a}
\begin{equation}
0=\mathcal{W}_{a}^{A}\left[\phi,\Delta\right]\equiv\Delta_{ab}^{-1}T_{bc}^{A}\phi_{c},
\end{equation}
and the normalization factor $N$ is chosen so that $\Gamma^{\mathrm{SI}}\left[\varphi,\Delta\right]$
numerically equals $\Gamma\left[\varphi,\Delta\right]$ when the arguments
satisfy the Ward identity \footnote{Formally $N=\left[\delta\left(0\right)\right]^{-1}$ though it is
not necessary to worry about rigorously defining this here. Also note
that if one wants invariance under redefinition of $\mathcal{W}$
then one also needs to insert a factor of $\mathrm{Det}\left(\frac{\delta\mathcal{W}}{\delta\phi}\right)$.
This can be handled by the introduction of Faddeev-Popov ghost fields.
However, this is not necessary here because we will not consider transformations
of $\mathcal{W}$.}. $T_{bc}^{A}$ is a generator of the $\mathrm{O}\left(N\right)$
symmetry where $A=1,\cdots,N\left(N-1\right)/2$ runs over the linearly
independent generators. When an explicit basis of generators is required
we take $T_{ab}^{jk}=i\left(\delta_{ja}\delta_{kb}-\delta_{jb}\delta_{ka}\right)$
where $A=\left(j,k\right)$ is thought of as an (antisymmetric) multi-index.
Note that the implicit integration convention can be maintained if
$T_{ab}^{A}\left(x,y\right)\propto\delta\left(x-y\right)$ contains
a spacetime delta function, though in this notation one must remember
that the upper indices do not have corresponding spacetime arguments
since they merely label the particular generator. $\Gamma^{\mathrm{SI}}\left[\varphi,\Delta\right]$
is defined only for field configurations satisfying the Ward identity,
and equals the usual effective action on those configurations. Thus
$\Gamma^{\mathrm{SI}}\left[\varphi,\Delta\right]$ is nothing but
the SI-2PIEA introduced by \citet{Pilaftsis2013}, arrived at in a
new way.

Proceeding from the hypothesis that the problems with symmetry improvement
are due to strict imposition of the constraint, as embodied by the
delta function above, we introduce a \emph{soft symmetry improved}
(SSI) effective action $\Gamma_{\xi}^{\mathrm{SSI}}\left[\varphi,\Delta\right]$
where the Ward identity is no longer strictly enforced. Small violations
$\mathcal{W}\neq0$ are allowed but punished in the functional integral.
A new free parameter controls how strictly the constraint is enforced.
The hope is that the added freedom allows consistent solutions with
non-trivial dynamics (e.g. linear response to external sources), while
the stiffness can be tuned to make violations of the Ward identity
acceptably small in practice. To achieve this, we replace the delta
function by a smoothed version $\delta\left(\mathcal{W}\right)\to\delta_{\xi}\left(\mathcal{W}\right)$
defined as follows
\begin{widetext}
\begin{align}
\exp\left(\frac{i}{\hbar}\Gamma_{\xi}^{\mathrm{SSI}}\left[\varphi,\Delta\right]\right) & =N_{0}\int\mathcal{D}\phi\ \delta\left(\phi-\varphi\right)\exp\left(\frac{i}{\hbar}\Gamma\left[\phi,\Delta\right]\right)\delta_{\xi}\left(\mathcal{W}\left[\phi,\Delta\right]\right)\nonumber \\
 & =N_{1}\int\mathcal{D}\left[\phi,\lambda_{\phi},\lambda_{W}\right]\ \exp\left(\frac{i}{\hbar}\left[\lambda_{\phi}\left(\phi-\varphi\right)+\Gamma\left[\phi,\Delta\right]+\lambda_{W}\mathcal{W}-\frac{1}{2}\xi\lambda_{W}^{2}\right]\right)\nonumber \\
 & =N_{2}\int\mathcal{D}\left[\phi,\lambda_{\phi}\right]\ \exp\left(\frac{i}{\hbar}\left[\lambda_{\phi}\left(\phi-\varphi\right)+\Gamma\left[\phi,\Delta\right]+\frac{1}{2\xi}\mathcal{W}^{2}\right]\right)\nonumber \\
 & =\exp\left(\frac{i}{\hbar}\left[\Gamma\left[\varphi,\Delta\right]+\frac{1}{2\xi}\mathcal{W}^{2}\left[\varphi,\Delta\right]\right]\right).
\end{align}
\end{widetext}

The first line is a formal expression that is defined by the next
line. The Fourier representation of the delta functions are used to
replace $\delta\left(\phi-\varphi\right)\to\int\mathcal{D}\lambda_{\phi}\exp\frac{i}{\hbar}\lambda_{\phi}\left(\phi-\varphi\right)$
etc. The $\frac{1}{2}\xi\lambda_{W}^{2}$ term is responsible for
smoothing the delta function, with the limit $\xi\to0$ corresponding
to a stiffening of the constraint. In the third line the integral
over $\lambda_{W}$, which is Gaussian, is performed. Finally, the
integral over $\lambda_{\phi}$ yields a delta function which kills
the $\phi$ integral, resulting in
\begin{equation}
\Gamma_{\xi}^{\mathrm{SSI}}\left[\varphi,\Delta\right]=\Gamma\left[\varphi,\Delta\right]+\frac{1}{2\xi}\mathcal{W}^{2}\left[\varphi,\Delta\right].
\end{equation}

The method can be generalized by using a weighted smoothing term $-\frac{1}{2}\xi\lambda_{W}R^{-1}\lambda_{W}$,
where $R^{-1}$ is an arbitrary positive definite symmetric kernel
which may depend on $\varphi$ and $\Delta$, which gives
\begin{equation}
\Gamma_{\xi R}^{\mathrm{SSI}}\left[\varphi,\Delta\right]=\Gamma\left[\varphi,\Delta\right]+\frac{1}{2\xi}\mathcal{W}R\mathcal{W}-\frac{i\hbar}{2}\mathrm{Tr}\ln R.
\end{equation}
The simpler form $\Gamma_{\xi}^{\mathrm{SSI}}\left[\varphi,\Delta\right]$
corresponds to a trivial kernel (now with indices explicit) 
\begin{equation}
R_{ab}^{AB}\left(x,y\right)=\delta^{AB}\delta_{ab}\delta\left(x-y\right),
\end{equation}
which is used exclusively in the following, though one should note
that the freedom to choose a non-trivial $R$ may be useful in certain
circumstances. The end result is simply that $\mathcal{W}=0$ is enforced
in the sense of (possibly weighted if $R$ is non-trivial) least-squared
error, rather than as a strict constraint.

We define the SSI equations of motion as the result of the variational
principle $\delta\Gamma_{\xi}^{\mathrm{SSI}}=0$, which gives:

\begin{align}
\frac{\delta\Gamma\left[\varphi,\Delta\right]}{\delta\varphi_{a}} & =-\frac{1}{\xi}\mathcal{W}_{c}^{A}\left[\varphi,\Delta\right]\frac{\delta}{\delta\varphi_{a}}\mathcal{W}_{c}^{A}\left[\varphi,\Delta\right]\nonumber \\
 & =-\frac{1}{\xi}\left(\Delta_{cf}^{-1}T_{fg}^{A}\varphi_{g}\right)\Delta_{cd}^{-1}T_{da}^{A},\\
\frac{\delta\Gamma\left[\varphi,\Delta\right]}{\delta\Delta_{ab}} & =-\frac{1}{\xi}\mathcal{W}_{c}^{A}\left[\varphi,\Delta\right]\frac{\delta}{\delta\Delta_{ab}}\mathcal{W}_{c}^{A}\left[\varphi,\Delta\right]\nonumber \\
 & =\frac{1}{\xi}\left(\Delta_{cf}^{-1}T_{fg}^{A}\varphi_{g}\right)\Delta_{ca}^{-1}\left(\Delta_{bd}^{-1}T_{de}^{A}\varphi_{e}\right).
\end{align}
Now the spontaneous symmetry breaking (SSB) ansatz

\begin{align}
\varphi_{a} & =v\delta_{aN},\\
\Delta_{ab}^{-1} & =\begin{cases}
\Delta_{G}^{-1} & a=b\neq N,\\
\Delta_{H}^{-1} & a=b=N,\\
0 & a\neq b
\end{cases}
\end{align}
can be used, where $\Delta_{G/H}$ are the Goldstone/Higgs propagators
respectively. This ansatz yields

\begin{align}
\frac{\delta\Gamma\left[\varphi,\Delta\right]}{\delta\varphi_{g}\left(x\right)} & =0,\ \left(g\neq N\right),\\
\frac{\delta\Gamma\left[\varphi,\Delta\right]}{\delta\varphi_{N}\left(x\right)} & =\frac{1}{\xi}2\left(N-1\right)v\int_{yz}\Delta_{G}^{-1}\left(y,z\right)\Delta_{G}^{-1}\left(y,x\right)\nonumber \\
 & =\frac{1}{\xi}2\left(N-1\right)vm_{G}^{4},
\end{align}

\begin{align}
\frac{\delta\Gamma\left[\varphi,\Delta\right]}{\delta\Delta_{G}\left(x,y\right)} & =-\frac{1}{\xi}2v^{2}\int_{wrz}\Delta_{G}^{-1}\left(w,r\right)\Delta_{G}^{-1}\left(w,x\right)\Delta_{G}^{-1}\left(y,z\right)\nonumber \\
 & =\frac{1}{\xi}2v^{2}m_{G}^{6},\\
\frac{\delta\Gamma\left[\varphi,\Delta\right]}{\delta\Delta_{H}} & =0,
\end{align}
where $m_{G}$ is the Goldstone mass.

Note that if one takes $\xi\to0$ proportionally to $vm_{G}^{4}$
one obtains for the non-trivial right hand sides above $2\left(N-1\right)vm_{G}^{4}/\xi\to\text{constant}$
and $2v^{2}m_{G}^{6}/\xi\to\left(\text{const.}\right)\times vm_{G}^{2}\to0$
and one recovers the usual SI-2PIEA scheme in the limit. In section
\ref{subsec:Broken-Phase-with-massless-Goldstones} this is shown
to hold with a careful treatment of the infinite volume limit. This
confirms the intuition that $\xi\to0$ approaches hard symmetry improvement
and that $\Gamma_{\xi}^{\mathrm{SSI}}\left[\varphi,\Delta\right]\to\Gamma^{\mathrm{SI}}\left[\varphi,\Delta\right]$
which really is just the standard symmetry improved effective action.
In the next sections these equations of motion are renormalized and
solved in the Hartree-Fock approximation.

\section{\label{sec:Renormalisation-of-HF}Renormalization of the Hartree-Fock
truncation}

There is a well established renormalization theory for 2PIEAs (see
e.g. \citep{Berges2005,Fejos2008,Fejos2014,VanHees2002}). Our renormalization
method is not particularly novel (we closely follow \citep{Pilaftsis2013,Brown2015a}),
but it is important to carefully treat the behavior of the theory
in the infrared which does lead to some new aspects. Therefore we
formulate the theory in Euclidean spacetime (i.e. the Matsubara formalism)
in a box of volume $\mathrm{V}=L^{3}$ with periodic boundary conditions
of period $L$ in the space directions and $\beta$ in the time $\tau=it$
direction. It turns out that the SSI method is sensitive to the manner
of taking the $\mathrm{V}\beta\to\infty$ limit. The Euclidean continuation
leads to $x=\left(t,\boldsymbol{x}\right)\to x_{E}\equiv\left(\tau,\boldsymbol{x}\right)$,
$\int_{x}\to-i\int_{x_{E}}$ and the conventions
\begin{align}
f\left(x_{E}\right) & =\frac{1}{\mathrm{V}\beta}\sum_{n,\boldsymbol{k}}\mathrm{e}^{i\left(\omega_{n}\tau+\boldsymbol{k}\cdot\boldsymbol{x}\right)}f\left(n,\boldsymbol{k}\right),\\
f\left(n,\boldsymbol{q}\right) & =\int_{x_{E}}\mathrm{e}^{-i\left(\omega_{m}\tau+\boldsymbol{q}\cdot\boldsymbol{x}\right)}f\left(x_{E}\right)
\end{align}
for Fourier transforms. The Matsubara frequencies are $\omega_{n}=2\pi n/\beta$
and the wave vectors $\boldsymbol{k}$ are discretized on a lattice
of spacing $2\pi/L$. The four dimensional Euclidean shorthand $k_{E}=\left(\omega_{n},\boldsymbol{k}\right)$
is often useful.

We will work in the Hartree-Fock approximation, which normally leads
a momentum independent self-energy and propagators of the form $\Delta_{G/H}^{-1}\left(n,\boldsymbol{k}\right)=k_{E}^{2}+m_{G/H}^{2}$.
However, it turns out that the SSI term leads to a momentum dependent
Goldstone self-energy. The equations of motion can be solved by treating
the Goldstone zero mode propagator $\Delta_{G}^{-1}\left(0,\boldsymbol{0}\right)$
as a dynamical variable separate from the non-zero modes. We \emph{define}
$m_{G}$ to be the mass associated with the \emph{non}zero Goldstone
modes, i.e. $\Delta_{G}^{-1}\left(n,\boldsymbol{k}\right)=k_{E}^{2}+m_{G}^{2}$
for $n,\boldsymbol{k}\neq0$. The zero mode propagator gains an independent
scaling factor $\Delta_{G}^{-1}\left(0,\boldsymbol{0}\right)\equiv\epsilon m_{G}^{2}$.
Note that the Goldstone theorem can be satisfied if $\epsilon=0$
even if $m_{G}^{2}\neq0$. This is the case for the novel $\beta\mathrm{V}\to\infty$
limit of the theory. The other limits are a reduction to the unimproved
2PIEA where $\epsilon=1$ and $m_{G}^{2}\neq0$ and a reduction to
the SI-2PIEA where $m_{G}^{2}=0$ and $\epsilon\neq0$.

The 2PIEA is derived from the partition functional
\begin{equation}
Z\left[J,K\right]=\int\mathcal{D}\left[\phi\right]\exp\left(-S_{E}\left[\phi\right]-J_{a}\phi_{a}-\frac{1}{2}\phi_{a}K_{ab}\phi_{b}\right),
\end{equation}
where
\begin{equation}
S_{E}\left[\phi\right]=\int_{x}\frac{1}{2}\left(\nabla\phi_{a}\right)^{2}+\frac{1}{2}m^{2}\phi_{a}\phi_{a}+\frac{1}{4!}\lambda\left(\phi_{a}\phi_{a}\right)^{2}
\end{equation}
is the Euclidean action. Then $W\left[J,K\right]=-\ln Z\left[J,K\right]$
is the connected generating functional and
\begin{equation}
\Gamma\left[\varphi,\Delta\right]=W-J\frac{\delta W}{\delta J}-K\frac{\delta W}{\delta K}\label{eq:legendre-transform}
\end{equation}
is the 2PIEA once $J$ and $K$ are eliminated in terms of $\varphi$
and $\Delta$ using
\begin{align}
\frac{\delta W}{\delta J_{a}} & =\left\langle \phi_{a}\right\rangle =\varphi_{a},\\
\frac{\delta W}{\delta K_{ab}} & =\frac{1}{2}\left\langle \phi_{a}\phi_{b}\right\rangle =\frac{1}{2}\left(\Delta_{ab}+\varphi_{a}\varphi_{b}\right).
\end{align}
The Legendre transform \eqref{eq:legendre-transform} can be evaluated
by the saddle point method, which results in the standard expression
\citep{Cornwall1974}
\begin{equation}
\Gamma=S_{E}\left[\varphi\right]+\frac{1}{2}\mathrm{Tr}\ln\left(\Delta^{-1}\right)+\frac{1}{2}\mathrm{Tr}\left(\Delta_{0}^{-1}\Delta-1\right)+\Gamma_{2},
\end{equation}
where $\Gamma_{2}$ is the set of two particle irreducible graphs
and

\begin{align}
\Delta_{0ab}^{-1} & \equiv\left.\frac{\delta^{2}S_{E}}{\delta\phi_{a}\delta\phi_{b}}\right|_{\phi\to\varphi}\nonumber \\
 & =\left(-\nabla^{2}+m^{2}+\frac{1}{6}\lambda\varphi^{2}\right)\delta_{ab}+\frac{1}{3}\lambda\varphi_{a}\varphi_{b}
\end{align}
is the unperturbed propagator. To $\mathcal{O}\left(\lambda\right)$,
\begin{eqnarray}
\Gamma_{2} & = & \frac{1}{4!}\lambda\Delta_{aa}\Delta_{bb}+\frac{1}{12}\lambda\Delta_{ab}\Delta_{ab}.
\end{eqnarray}

To form $\Gamma_{\xi}^{\mathrm{SSI}}$ one adds the soft-symmetry
improvement term $-\frac{1}{2\xi}\mathcal{W}^{2}$. Note that $\mathcal{W}$
is pure imaginary due to the $i$ in $T^{A}$ so $-\mathcal{W}^{2}$
is positive definite. After using the SSB ansatz the condition $\mathcal{W}_{a}^{A}=0$
becomes Goldstone's theorem $v\Delta_{G}^{-1}\left(0,\boldsymbol{0}\right)=0$.
We drop an irrelevant constant, use the SSB ansatz and insert the
renormalization constants $Z$, $Z_{\Delta}$, $\delta m_{0,1}^{2}$,
$\delta\lambda_{0}$, and $\delta\lambda_{1,2}^{A,B}$ by making the
replacements (c.f. \citep{Brown2015a})

\begin{eqnarray}
\left(\phi,\varphi,v\right) & \to & Z^{1/2}\left(\phi,\varphi,v\right),\\
m^{2} & \to & Z^{-1}Z_{\Delta}^{-1}\left(m^{2}+\delta m^{2}\right),\\
\lambda & \to & Z^{-2}\left(\lambda+\delta\lambda\right),\\
\Delta & \to & ZZ_{\Delta}\Delta.
\end{eqnarray}
Due to the presence of composite operators in the effective action,
additional renormalization constants are required compared to the
standard perturbative renormalization theory: $\delta m_{0}^{2}$
and $\delta\lambda_{0}$ for terms in the bare action, $\delta m_{1}^{2}$
for one-loop terms, $\delta\lambda_{1}^{A}$ for terms of the form
$\phi_{a}\phi_{a}\Delta_{bb}$, $\delta\lambda_{1}^{B}$ for $\phi_{a}\phi_{b}\Delta_{ab}$
terms, $\delta\lambda_{2}^{A}$ for $\Delta_{aa}\Delta_{bb}$ and
$\delta\lambda_{2}^{B}$ for $\Delta_{ab}\Delta_{ab}$. The fact that
extra counterterms are required to renormalize the 2PIEA is not a
problem so long as a sufficient number of renormalization conditions
can be found. Altogether there are nine renormalization constants
which must be eliminated by imposing nine conditions. It turns out
that $Z=Z_{\Delta}=1$ in the Hartree-Fock approximation due to the
momentum independence of the UV divergences in this approximation
(this is well known in the 2PIEA literature \citep{Berges2005,Fejos2008},
but it may not be immediately obvious that it continues to hold with
the addition of the SSI terms; indeed it does). One can introduce
a renormalization constant $Z_{\xi}$ for $\xi$ but this turns out
to be unnecessary. Thus we arrive at the renormalized SSI effective
action:
\begin{widetext}
\begin{eqnarray}
\Gamma_{\xi}^{\mathrm{SSI}}\left[\varphi,\Delta\right] & = & \int_{x}\left(\frac{m^{2}+\delta m_{0}^{2}}{2}v^{2}+\frac{\lambda+\delta\lambda_{0}}{4!}v^{4}\right)+\frac{1}{2}\left(N-1\right)\mathrm{Tr}\ln\left(\Delta_{G}^{-1}\right)+\frac{1}{2}\mathrm{Tr}\ln\left(\Delta_{H}^{-1}\right)\nonumber \\
 &  & +\frac{1}{2}\left(N-1\right)\mathrm{Tr}\left[\left(-\nabla^{2}+m^{2}+\delta m_{1}^{2}+\frac{\lambda+\delta\lambda_{1}^{A}}{6}v^{2}\right)\Delta_{G}\right]+\frac{1}{2}\mathrm{Tr}\left[\left(-\nabla^{2}+m^{2}+\delta m_{1}^{2}\vphantom{\frac{3\lambda+\delta\lambda_{1}^{A}+2\delta\lambda_{1}^{B}}{6}}\right.\right.\nonumber \\
 &  & \left.\left.+\frac{3\lambda+\delta\lambda_{1}^{A}+2\delta\lambda_{1}^{B}}{6}v^{2}\right)\Delta_{H}\right]+\Gamma_{2}+\frac{1}{\xi}\left(N-1\right)v^{2}\int_{xyz}\Delta_{G}^{-1}\left(x,y\right)\Delta_{G}^{-1}\left(x,z\right)
\end{eqnarray}
\end{widetext}

with

\begin{eqnarray}
\Gamma_{2} & = & \frac{1}{4!}\left(N-1\right)\nonumber \\
 &  & \times\left[\left(N+1\right)\lambda+\left(N-1\right)\delta\lambda_{2}^{A}+2\delta\lambda_{2}^{B}\right]\Delta_{G}\Delta_{G}\nonumber \\
 &  & +\frac{1}{4!}\left(\lambda+\delta\lambda_{2}^{A}\right)2\left(N-1\right)\Delta_{G}\Delta_{H}\nonumber \\
 &  & +\frac{1}{4!}\left(3\lambda+\delta\lambda_{2}^{A}+2\delta\lambda_{2}^{B}\right)\Delta_{H}\Delta_{H}\nonumber \\
 &  & +\mathcal{O}\left(\lambda^{2}\right).
\end{eqnarray}

$\Gamma_{\xi}^{\mathrm{SSI}}\left[\varphi,\Delta\right]$ can be simplified
using the mode expansions

\begin{equation}
\Delta_{G/H}\left(x_{E},y_{E}\right)=\frac{1}{\mathrm{V}\beta}\sum_{n,\boldsymbol{k}}\mathrm{e}^{ik_{E}\cdot\left(x_{E}-y_{E}\right)}\Delta_{G/H}\left(n,\boldsymbol{k}\right)
\end{equation}
and doing the integrals, noting that the integrals in the SSI term
give

\begin{equation}
\int_{xyz}\Delta_{G}^{-1}\left(x,y\right)\Delta_{G}^{-1}\left(x,z\right)=\mathrm{V}\beta\left[\Delta_{G}^{-1}\left(0,\boldsymbol{0}\right)\right]^{2}.
\end{equation}
The result is

\begin{eqnarray}
\Gamma_{\xi}^{\mathrm{SSI}}\left[\varphi,\Delta\right] & = & \mathrm{V}\beta\left(\frac{m^{2}+\delta m_{0}^{2}}{2}v^{2}+\frac{\lambda+\delta\lambda_{0}}{4!}v^{4}\right)\nonumber \\
 &  & +\frac{1}{2}\left(N-1\right)\sum_{n,\boldsymbol{k}}\ln\frac{1}{\Delta_{G}\left(n,\boldsymbol{k}\right)}\nonumber \\
 &  & +\frac{1}{2}\sum_{n,\boldsymbol{k}}\ln\frac{1}{\Delta_{H}\left(n,\boldsymbol{k}\right)}\nonumber \\
 &  & +\frac{1}{2}\left(N-1\right)\sum_{n,\boldsymbol{k}}\left(k_{E}^{2}+m^{2}+\delta m_{1}^{2}\right.\nonumber \\
 &  & \left.+\frac{\lambda+\delta\lambda_{1}^{A}}{6}v^{2}\right)\Delta_{G}\left(n,\boldsymbol{k}\right)\nonumber \\
 &  & +\frac{1}{2}\sum_{n,\boldsymbol{k}}\left(k_{E}^{2}+m^{2}+\delta m_{1}^{2}\right.\nonumber \\
 &  & \left.+\frac{3\lambda+\delta\lambda_{1}^{A}+2\delta\lambda_{1}^{B}}{6}v^{2}\right)\Delta_{H}\left(n,\boldsymbol{k}\right)+\Gamma_{2}\nonumber \\
 &  & +\frac{1}{\xi}\left(N-1\right)v^{2}\mathrm{V}\beta\left[\Delta_{G}^{-1}\left(0,\boldsymbol{0}\right)\right]^{2}.
\end{eqnarray}

As a brief digression a simple consistency check can be performed
by examining the tree level equations of motion, which are (setting
renormalization constants to their trivial values)
\begin{equation}
0=v\left\{ \left(m^{2}+\frac{\lambda}{6}v^{2}\right)+\frac{2\left(N-1\right)}{\xi}\left[\Delta_{G}^{-1}\left(0,\boldsymbol{0}\right)\right]^{2}\right\} ,\label{eq:tree-level-vev-eom}
\end{equation}

\begin{align}
\Delta_{G}^{-1}\left(n,\boldsymbol{k}\right) & =k_{E}^{2}+m^{2}+\frac{\lambda}{6}v^{2},\ n,\boldsymbol{k}\neq0,\\
\Delta_{G}^{-1}\left(0,\boldsymbol{0}\right) & =m^{2}+\frac{\lambda}{6}v^{2}-\frac{4\mathrm{V}\beta}{\xi}v^{2}\left[\Delta_{G}^{-1}\left(0,\boldsymbol{0}\right)\right]^{3},\label{eq:tree-level-zero-mode-eom}\\
\Delta_{H}^{-1}\left(n,\boldsymbol{k}\right) & =k_{E}^{2}+m^{2}+\frac{\lambda}{2}v^{2}.
\end{align}
Indeed the classical solution $v^{2}=-6m^{2}/\lambda$, $\Delta_{G}^{-1}\left(n,\boldsymbol{k}\right)=k_{E}^{2}$
and $\Delta_{H}^{-1}\left(n,\boldsymbol{k}\right)=k_{E}^{2}+m_{H}^{2}=k_{E}^{2}+\frac{\lambda}{3}v^{2}$
is consistent with these as expected. However, since these equations
are self-consistent, spurious solutions are also possible. This can
be investigated by solving \eqref{eq:tree-level-vev-eom} and \eqref{eq:tree-level-zero-mode-eom}
together on the assumption that $v^{2}\neq0,-6m^{2}/\lambda$. Using
\eqref{eq:tree-level-vev-eom} to reduce the degree of \eqref{eq:tree-level-zero-mode-eom}
to first order gives the potentially spurious solution

\begin{align}
-\frac{\xi}{2\left(N-1\right)} & =\left(m^{2}+\frac{\lambda}{6}v^{2}\right)/\left[1-\frac{2\mathrm{V}\beta}{N-1}v^{2}\left(m^{2}+\frac{\lambda}{6}v^{2}\right)\right]^{2},\label{eq:spurious-soln-chi-eq}\\
\Delta_{G}^{-1}\left(0,\boldsymbol{0}\right) & =\left(m^{2}+\frac{\lambda}{6}v^{2}\right)/\left[1-\frac{2\mathrm{V}\beta}{N-1}v^{2}\left(m^{2}+\frac{\lambda}{6}v^{2}\right)\right].
\end{align}
The condition that there are no tachyons requires $\Delta_{G}^{-1}\left(0,\boldsymbol{0}\right)\geq0$
which implies

\begin{equation}
0\leq m^{2}+\frac{\lambda}{6}v^{2}<\frac{N-1}{2\mathrm{V}\beta v^{2}}.
\end{equation}
This then implies that the right hand side of \eqref{eq:spurious-soln-chi-eq}
is positive, but then the left hand side $\propto-\xi$ is negative,
leading to a contradiction. Thus the only spurious solutions are tachyonic
and so easily dismissable.

Returning to the main line of the argument, the rest of the paper
restricts attention to the Hartree-Fock truncation where only the
$\mathcal{O}\left(\lambda\right)$ terms in $\Gamma_{2}$ are kept.
Thus
\begin{widetext}
\begin{eqnarray}
\Gamma_{2} & = & \frac{1}{4!}\left(N-1\right)\left[\left(N+1\right)\lambda+\left(N-1\right)\delta\lambda_{2}^{A}+2\delta\lambda_{2}^{B}\right]\frac{1}{\mathrm{V}\beta}\sum_{n,\boldsymbol{k}}\Delta_{G}\left(n,\boldsymbol{k}\right)\sum_{j,\mathbf{q}}\Delta_{G}\left(j,\mathbf{q}\right)\nonumber \\
 &  & +\frac{1}{4!}\left(\lambda+\delta\lambda_{2}^{A}\right)2\left(N-1\right)\frac{1}{\mathrm{V}\beta}\sum_{n,\boldsymbol{k}}\Delta_{G}\left(n,\boldsymbol{k}\right)\sum_{j,\mathbf{q}}\Delta_{H}\left(j,\mathbf{q}\right)\nonumber \\
 &  & +\frac{1}{4!}\left(3\lambda+\delta\lambda_{2}^{A}+2\delta\lambda_{2}^{B}\right)\frac{1}{\mathrm{V}\beta}\sum_{n,\boldsymbol{k}}\Delta_{H}\left(n,\boldsymbol{k}\right)\sum_{j,\mathbf{q}}\Delta_{H}\left(j,\mathbf{q}\right).
\end{eqnarray}
The resulting equations of motion are the vev equation

\begin{eqnarray}
0 & = & \mathrm{V}\beta\left(\frac{m^{2}+\delta m_{0}^{2}}{2}2v+\frac{\lambda+\delta\lambda_{0}}{4!}4v^{3}\right)+\left(N-1\right)\frac{\lambda+\delta\lambda_{1}^{A}}{6}v\sum_{n,\boldsymbol{k}}\Delta_{G}\left(n,\boldsymbol{k}\right)\nonumber \\
 &  & +\frac{3\lambda+\delta\lambda_{1}^{A}+2\delta\lambda_{1}^{B}}{6}v\sum_{n,\boldsymbol{k}}\Delta_{H}\left(n,\boldsymbol{k}\right)+\frac{1}{\xi}\left(N-1\right)2v\mathrm{V}\beta\left[\Delta_{G}^{-1}\left(0,\boldsymbol{0}\right)\right]^{2},
\end{eqnarray}
\end{widetext}

the Goldstone propagator equation

\begin{eqnarray}
\frac{1}{\Delta_{G}\left(n,\boldsymbol{k}\right)} & = & k_{E}^{2}+m^{2}+\delta m_{1}^{2}+\frac{\lambda+\delta\lambda_{1}^{A}}{6}v^{2}\nonumber \\
 &  & +\frac{1}{6}\left[\left(N+1\right)\lambda+\left(N-1\right)\delta\lambda_{2}^{A}+2\delta\lambda_{2}^{B}\right]\nonumber \\
 &  & \times\frac{1}{\mathrm{V}\beta}\sum_{j,\mathbf{q}}\Delta_{G}\left(j,\mathbf{q}\right)\nonumber \\
 &  & +\frac{1}{3!}\left(\lambda+\delta\lambda_{2}^{A}\right)\frac{1}{\mathrm{V}\beta}\sum_{j,\mathbf{q}}\Delta_{H}\left(j,\mathbf{q}\right)\nonumber \\
 &  & -\delta_{n0}\delta_{\boldsymbol{k}\boldsymbol{0}}\frac{4v^{2}}{\xi}\mathrm{V}\beta\left[\Delta_{G}^{-1}\left(0,\boldsymbol{0}\right)\right]^{3},
\end{eqnarray}
and the Higgs propagator equation

\begin{eqnarray}
\frac{1}{\Delta_{H}\left(n,\boldsymbol{k}\right)} & = & k_{E}^{2}+m^{2}+\delta m_{1}^{2}\nonumber \\
 &  & +\frac{3\lambda+\delta\lambda_{1}^{A}+2\delta\lambda_{1}^{B}}{6}v^{2}\nonumber \\
 &  & +\frac{1}{3!}\left(\lambda+\delta\lambda_{2}^{A}\right)\left(N-1\right)\frac{1}{\mathrm{V}\beta}\sum_{j,\mathbf{q}}\Delta_{G}\left(j,\mathbf{q}\right)\nonumber \\
 &  & +\frac{1}{3!}\frac{3\lambda+\delta\lambda_{2}^{A}+2\delta\lambda_{2}^{B}}{\mathrm{V}\beta}\sum_{j,\mathbf{q}}\Delta_{H}\left(j,\mathbf{q}\right).
\end{eqnarray}

As previously mentioned, the self-energies are momentum independent
except for the $\delta_{n0}\delta_{\boldsymbol{k}\boldsymbol{0}}$
term in $\Delta_{G}$. Therefore we write the propagators as
\begin{eqnarray}
\Delta_{G}\left(n,\boldsymbol{k}\right) & = & \begin{cases}
\Delta_{G}\left(0,\boldsymbol{0}\right) & n=\boldsymbol{k}=0\\
\frac{1}{k_{E}^{2}+m_{G}^{2}} & n,\boldsymbol{k}\neq0
\end{cases},\\
\Delta_{H}\left(n,\boldsymbol{k}\right) & = & \frac{1}{k_{E}^{2}+m_{H}^{2}},
\end{eqnarray}
and define $\Delta_{G}^{-1}\left(0,\boldsymbol{0}\right)\equiv\epsilon m_{G}^{2}$
which is now independent of the nonzero modes. The zero mode obeys
the equation

\begin{equation}
\Delta_{G}^{-1}\left(0,\boldsymbol{0}\right)=m_{G}^{2}-4\frac{1}{\xi}v^{2}\mathrm{V}\beta\left[\Delta_{G}^{-1}\left(0,\boldsymbol{0}\right)\right]^{3}.
\end{equation}

Now there are two cases which must be distinguished. In the first,
$m_{G}^{2}=0$ and the zero mode equation has the solutions

\begin{align}
\Delta_{G}^{-1}\left(0,\boldsymbol{0}\right) & =\begin{cases}
0,\\
\pm i\sqrt{\frac{\xi}{4v^{2}\mathrm{V}\beta}},
\end{cases}
\end{align}
the latter two of which are clearly unphysical. However the first
solution is just what one would have if $\Delta_{G}\left(n,\boldsymbol{k}\right)=k_{E}^{-2}$
as usual for a massless particle (i.e. the zero mode need no longer
be treated separately). Then $\sum_{j,\mathbf{q}}\Delta_{G}\left(j,\mathbf{q}\right)$
and $\sum_{j,\mathbf{q}}\Delta_{H}\left(j,\mathbf{q}\right)$ are
just the familiar Hartree-Fock tadpole sums, which in the infinite
volume limit are
\begin{eqnarray}
\sum_{n,\boldsymbol{k}}\Delta_{G}\left(n,\boldsymbol{k}\right) & = & \beta\mathrm{V}\left(\mathcal{T}_{G}^{\infty}+\mathcal{T}_{G}^{\mathrm{fin}}+\mathcal{T}_{G}^{\mathrm{th}}\right),\label{eq:goldstone-tadpole-sum}\\
\sum_{n,\boldsymbol{k}}\Delta_{H}\left(n,\boldsymbol{k}\right) & = & \beta\mathrm{V}\left(\mathcal{T}_{H}^{\infty}+\mathcal{T}_{H}^{\mathrm{fin}}+\mathcal{T}_{H}^{\mathrm{th}}\right),\label{eq:higgs-tadpole-sum}
\end{eqnarray}
where

\begin{eqnarray}
\mathcal{T}_{G/H}^{\infty} & = & -\frac{m_{G/H}^{2}}{16\pi^{2}}\left[\frac{1}{\eta}-\gamma+1+\ln\left(4\pi\right)\right],\\
\mathcal{T}_{G/H}^{\mathrm{fin}} & = & \frac{m_{G/H}^{2}}{16\pi^{2}}\ln\frac{m_{G/H}^{2}}{\mu^{2}}
\end{eqnarray}
are the vacuum contributions in dimensional regularization in $4-2\eta$
dimensions with $\overline{\mathrm{MS}}$ subtraction at the scale
$\mu$ ($\gamma\approx0.577$ is the Euler gamma constant) and $\mathcal{T}_{G/H}^{\mathrm{th}}$
are the Bose-Einstein integrals 
\begin{equation}
\mathcal{T}_{G/H}^{\mathrm{th}}=\int_{\boldsymbol{k}}\frac{1}{\omega_{\boldsymbol{k}}}\frac{1}{\mathrm{e}^{\beta\omega_{\boldsymbol{k}}}-1},\ \omega_{\boldsymbol{k}}=\sqrt{\boldsymbol{k}^{2}+m_{G/H}^{2}}.
\end{equation}

\begin{figure}
\includegraphics[width=1\columnwidth]{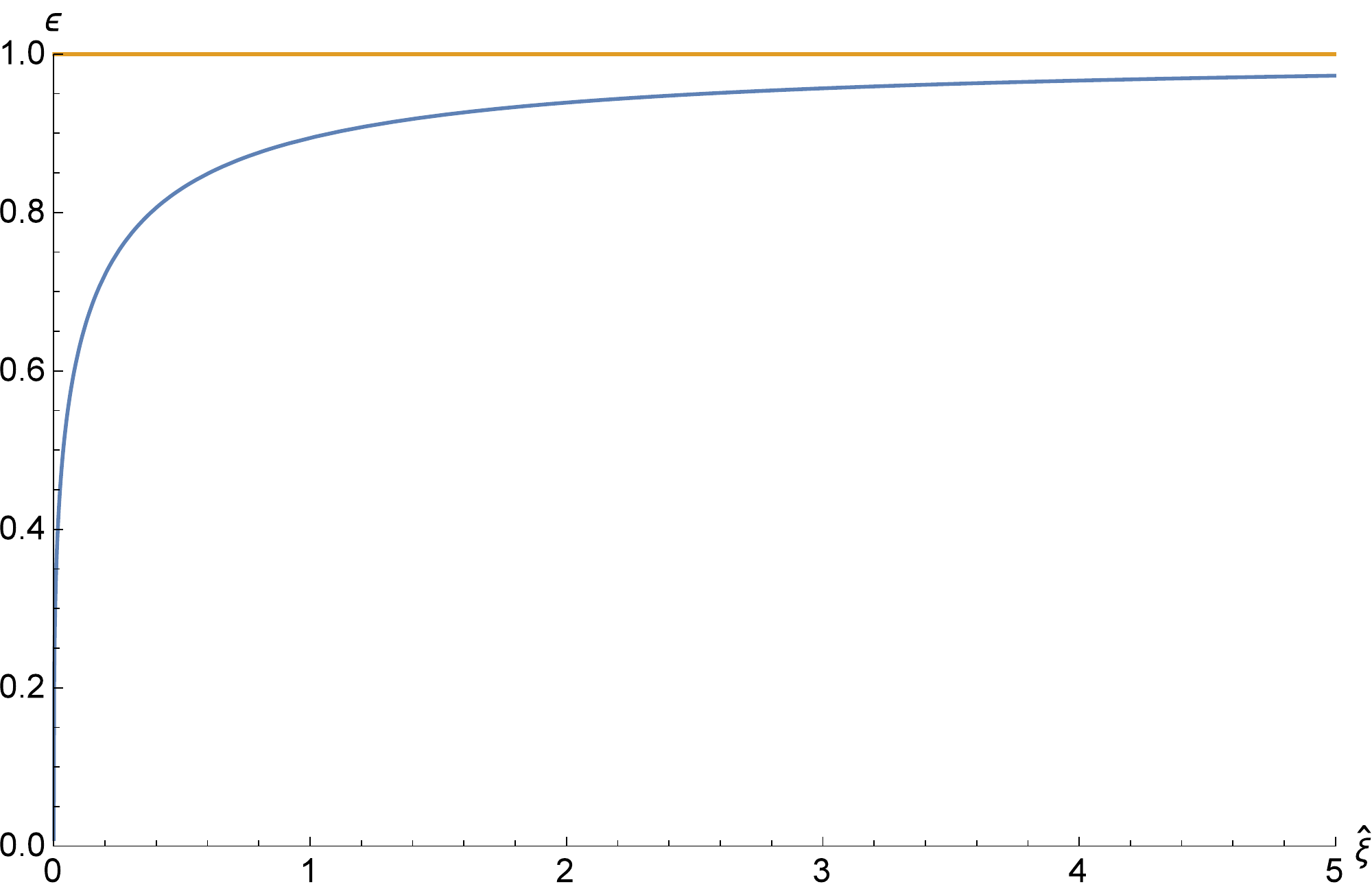}

\caption{\label{fig:goldstone-zero-mode-epsilon}(Color online) Plot of $\epsilon$
(blue) versus $\hat{\xi}$ solving the Goldstone boson zero mode equation
\eqref{eq:epsilon-solution}. The line at $\epsilon=1$ is to guide
the eye.}
\end{figure}

If, on the other hand, $m_{G}^{2}\neq0$ then $\Delta_{G}$ no longer
has the usual form and the Goldstone tadpole must be handled differently.
In this case it can be rewritten as
\begin{eqnarray}
\sum_{j,\mathbf{q}}\Delta_{G}\left(j,\mathbf{q}\right) & = & \sum_{j,\mathbf{q}\neq0}\Delta_{G}\left(j,\mathbf{q}\right)+\Delta_{G}\left(0,\boldsymbol{0}\right)\nonumber \\
 & = & \sum_{j,\mathbf{q}\neq0}\Delta_{G}\left(j,\mathbf{q}\right)+\frac{1}{m_{G}^{2}}\nonumber \\
 &  & +\Delta_{G}\left(0,\boldsymbol{0}\right)-\frac{1}{m_{G}^{2}}\nonumber \\
 & = & \sum_{j,\mathbf{q}}\tilde{\Delta}_{G}\left(j,\mathbf{q}\right)\nonumber \\
 &  & +\Delta_{G}\left(0,\boldsymbol{0}\right)-\frac{1}{m_{G}^{2}},
\end{eqnarray}
where $\tilde{\Delta}_{G}$ is an auxiliary propagator defined to
have the usual form
\begin{equation}
\tilde{\Delta}_{G}\left(n,\boldsymbol{k}\right)=\frac{1}{k_{E}^{2}+m_{G}^{2}}.
\end{equation}
Then $\sum_{j,\mathbf{q}}\tilde{\Delta}_{G}\left(j,\mathbf{q}\right)$
is just the familiar Hartree-Fock tadpole sum for a particle of mass
$m_{G}$. The terms $\Delta_{G}\left(0,\boldsymbol{0}\right)-\frac{1}{m_{G}^{2}}$
in the Goldstone tadpole account for the shift in the zero mode propagator
from its usual value. The zero mode equation can be rewritten as

\begin{align}
\epsilon & =1-\frac{4v^{2}\mathrm{V}\beta m_{G}^{4}}{\xi}\epsilon^{3}=1-\frac{4\epsilon^{3}}{27\hat{\xi}},
\end{align}
where
\begin{equation}
\hat{\xi}=\frac{\xi}{27v^{2}\mathrm{V}\beta m_{G}^{4}}
\end{equation}
is dimensionless and the numeric factor has been chosen for later
convenience. The real solution of this cubic equation is
\begin{equation}
\epsilon=\frac{3}{2}\sqrt[3]{\hat{\xi}\left(\sqrt{\hat{\xi}+1}-1\right)}\left(\sqrt[3]{\frac{\hat{\xi}}{\left(\sqrt{\hat{\xi}+1}-1\right)^{2}}}-1\right),\label{eq:epsilon-solution}
\end{equation}
which is monotonically increasing from $0$ to $1$ as $\hat{\xi}$
goes from $0$ to $+\infty$ and behaves asymptotically as
\begin{equation}
\epsilon\sim\begin{cases}
\frac{3\hat{\xi}^{1/3}}{2^{2/3}}+\mathcal{O}\left(\hat{\xi}^{2/3}\right), & \hat{\xi}\to0,\\
1-\frac{4}{27\hat{\xi}}+\mathcal{O}\left(\hat{\xi}^{-2}\right), & \hat{\xi}\to\infty.
\end{cases}\label{eq:epsilon-asymptotic-behaviour}
\end{equation}
The behavior of $\epsilon$ is shown in Fig. \ref{fig:goldstone-zero-mode-epsilon}.

The remaining equations are renormalized by demanding that kinematically
distinct divergences vanish, essentially duplicating the renormalization
method of \citep{Pilaftsis2013,Brown2015a}. This is done by substituting
the tadpoles \eqref{eq:goldstone-tadpole-sum}-\eqref{eq:higgs-tadpole-sum}
in the equations of motion, rearranging to obtain expressions for
$v$, $m_{G}^{2}$ and $m_{H}^{2}$, then demanding that the divergences
proportional to $v$, $\mathcal{T}_{G}^{\mathrm{fin}}$ and $\mathcal{T}_{H}^{\mathrm{fin}}$
independently vanish. This leads to eleven equations which are nontrivially
consistent and determine the nine counterterms. No new difficulties
are found here compared to the standard treatment and the details
are left in a supplemental \noun{Mathematica} notebook \citep{Supp}.
The resulting finite equations of motion are the vev equation

\begin{eqnarray}
v & = & 0,\\
 & \text{or}\nonumber \\
0 & = & m^{2}+\frac{\lambda}{6}v^{2}+\left(N-1\right)\frac{\lambda}{6}\nonumber \\
 &  & \times\left[\frac{m_{G}^{2}}{16\pi^{2}}\ln\frac{m_{G}^{2}}{\mu^{2}}+\mathcal{T}_{G}^{\mathrm{th}}+\frac{1}{\mathrm{V}\beta}\frac{1}{m_{G}^{2}}\left(\frac{1}{\epsilon}-1\right)\right]\nonumber \\
 &  & +\frac{\lambda}{2}\left(\frac{m_{H}^{2}}{16\pi^{2}}\ln\frac{m_{H}^{2}}{\mu^{2}}+\mathcal{T}_{H}^{\mathrm{th}}\right)\nonumber \\
 &  & +\frac{\left(N-1\right)2}{\xi}\left(m_{G}^{2}\epsilon\right)^{2},\label{eq:SSI-2PI-HF-vev-eom}
\end{eqnarray}
the Goldstone gap equation

\begin{eqnarray}
m_{G}^{2} & = & m^{2}+\frac{\lambda}{6}v^{2}+\left(N+1\right)\frac{\lambda}{6}\nonumber \\
 &  & \times\left[\frac{m_{G}^{2}}{16\pi^{2}}\ln\frac{m_{G}^{2}}{\mu^{2}}+\mathcal{T}_{G}^{\mathrm{th}}+\frac{1}{\mathrm{V}\beta}\frac{1}{m_{G}^{2}}\left(\frac{1}{\epsilon}-1\right)\right]\nonumber \\
 &  & +\frac{\lambda}{6}\left(\frac{m_{H}^{2}}{16\pi^{2}}\ln\frac{m_{H}^{2}}{\mu^{2}}+\mathcal{T}_{H}^{\mathrm{th}}\right),\label{eq:SSI-2PI-HF-G-eom}
\end{eqnarray}
and the Higgs gap equation

\begin{eqnarray}
m_{H}^{2} & = & m^{2}+\frac{\lambda}{2}v^{2}+\left(N-1\right)\frac{\lambda}{6}\nonumber \\
 &  & \times\left[\frac{m_{G}^{2}}{16\pi^{2}}\ln\frac{m_{G}^{2}}{\mu^{2}}+\mathcal{T}_{G}^{\mathrm{th}}+\frac{1}{\mathrm{V}\beta}\frac{1}{m_{G}^{2}}\left(\frac{1}{\epsilon}-1\right)\right]\nonumber \\
 &  & +\frac{\lambda}{2}\left(\frac{m_{H}^{2}}{16\pi^{2}}\ln\frac{m_{H}^{2}}{\mu^{2}}+\mathcal{T}_{H}^{\mathrm{th}}\right).\label{eq:SSI-2PI-HF-H-eom}
\end{eqnarray}

\section{\label{sec:Solution-in-the-infinite-volume-limit}Solution in the
infinite volume / low temperature limit}

\subsection{Scaling of the solutions\label{subsec:Scaling-of-the-solutions}}

We desire solutions of \eqref{eq:SSI-2PI-HF-vev-eom}-\eqref{eq:SSI-2PI-HF-H-eom}
in the $\mathrm{V}\beta\to\infty$ limit. It is possible in general
to look for solutions in the symmetric and broken phases with scalings
$\xi\sim\left(\mathrm{V}\beta\right)^{\alpha}$ and $m_{G}^{2}\sim\left(\mathrm{V}\beta\right)^{-\gamma}$
for $\gamma\geq0$. In section \ref{subsec:Symmetric-Phase} we examine
the symmetric phase ($v=0$, $m_{G}\neq0$) and show that it is unaffected
by SSI as expected. In section \ref{subsec:Broken-Phase-with-massive-Goldstones}
we examine the broken phase $v\neq0$ with $m_{G}\neq0$. Ordinarily
Goldstone's theorem is broken in this regime, however with the additional
freedom afforded by SSI we find a scaling for $\xi$ such that $\epsilon\to0$
and Goldstone's theorem is nevertheless satisfied. Finally, in section
\ref{subsec:Broken-Phase-with-massless-Goldstones} we examine the
broken phase $v\neq0$ with massless Goldstones $m_{G}=0$. We expect
SSI in this regime to be close to the old symmetry improvement method
and in fact it turns out to be exactly equivalent. The apparent extra
freedom of the SSI method (the choice of $\xi$) is equivalent the
freedom to choose the Lagrange multiplier of the SI method, which
we demonstrate by deriving the explicit connection between them. This
gives a new insight into why the SI equations of motion do not depend
on the Lagrange multiplier, which previously appeared as a remarkable
coincidence but now can be understood as a consequence of the $\mathrm{V}\beta\to\infty$
limit.

The effects of SSI enter into \eqref{eq:SSI-2PI-HF-vev-eom}-\eqref{eq:SSI-2PI-HF-H-eom}
through two terms, for which we introduce the shorthands
\begin{eqnarray}
\mathcal{S}_{1} & = & \frac{1}{\mathrm{V}\beta}\frac{1}{m_{G}^{2}}\left(\frac{1}{\epsilon}-1\right),\\
\mathcal{S}_{2} & = & \frac{2\left(N-1\right)}{\xi}\left(m_{G}^{2}\epsilon\right)^{2}.
\end{eqnarray}
In the following sections we consider all possible scalings of $\xi$
and $m_{G}$ and their effect on these terms, ruling out most possibilities.
For reference purposes we collect here the scalings that work in each
section. In section \ref{subsec:Symmetric-Phase} we find that the
symmetric phase exists independent of the $\mathrm{V}\beta\to\infty$
limit. In section \ref{subsec:Broken-Phase-with-massive-Goldstones}
it is necessary to let $\xi$ scale as $\xi=\left(\mathrm{V}\beta\right)^{-2}\zeta$
where $\zeta$ is a constant (with mass dimension $\left[\zeta\right]=-6$),
for which

\begin{eqnarray}
\epsilon & \to & \frac{1}{\mathrm{V}\beta}\left(\frac{\zeta}{4v^{2}m_{G}^{4}}\right)^{1/3}\to0,\\
\mathcal{S}_{1} & \to & \left(\frac{4v^{2}}{\zeta m_{G}^{2}}\right)^{1/3},\\
\mathcal{S}_{2} & \to & \frac{1}{\zeta}\left(N-1\right)\left(\frac{\zeta m_{G}^{2}}{\sqrt{2}v^{2}}\right)^{2/3}.
\end{eqnarray}
The equations of motion are then nondimensionalized and studied using
three methods: perturbation theory in $\zeta^{-1}$, at leading order
in $1/N$ and through exact numerical solutions. Finally, in section
\ref{subsec:Broken-Phase-with-massless-Goldstones} both $\xi$ and
$m_{G}$ must be scaled as

\begin{eqnarray}
\xi & = & \left(\mathrm{V}\beta\right)^{\alpha}\mu^{2+4\alpha}\hat{\zeta},\\
m_{G}^{2} & = & \left(\mathrm{V}\beta\right)^{-\gamma}\mu^{2-4\gamma}y,
\end{eqnarray}
where $\gamma>0$, $\alpha+2\gamma+2=0$ and $\hat{\zeta}$ and $y$
are dimensionless. Any scaling satisfying these conditions leads to
identical equations of motion and solutions. Then

\begin{eqnarray}
\epsilon & \to & \frac{1}{\mathrm{V}\beta\mu^{4}}\left(\frac{\hat{\zeta}}{4xy^{2}}\right)^{1/3},\\
\mathcal{S}_{1} & \to & 0,\\
\mathcal{S}_{2} & \to & \mu^{2}\left(N-1\right)\left(\frac{y^{2}}{2\hat{\zeta}x^{2}}\right)^{1/3},
\end{eqnarray}
where $x=v^{2}/\mu^{2}$ is the dimensionless vev.

\subsection{Symmetric phase\label{subsec:Symmetric-Phase}}

At high temperatures there should be a symmetric phase solution to
the equations of motion. We therefore examine the $v\to0$ limit of
the equations of motion. As $v\to0$ at fixed $\xi$, $\epsilon\to1-\frac{4v^{2}\mathrm{V}\beta m_{G}^{4}}{\xi}$
provided $m_{G}$ does not go to infinity faster than $1/\sqrt{v}$.
Then
\begin{equation}
\mathcal{S}_{1}\to\frac{4v^{2}m_{G}^{2}}{\xi}\to0,
\end{equation}
and the equations of motion \eqref{eq:SSI-2PI-HF-vev-eom}-\eqref{eq:SSI-2PI-HF-H-eom}
reduce to

\begin{equation}
m_{G}^{2}=m_{H}^{2}=m^{2}+\frac{1}{6}\left(N+2\right)\lambda\left(\frac{m_{H}^{2}}{16\pi^{2}}\ln\frac{m_{H}^{2}}{\mu^{2}}+\mathcal{T}_{H}^{\mathrm{th}}\right),
\end{equation}
which is a symmetric (i.e. equal mass) phase as expected. Indeed,
the gap equation is unmodified by SSI in the symmetric phase because
there $\mathcal{W}=0$ trivially. This phase terminates at the critical
point $m_{G}^{2}=m_{H}^{2}=0$ which gives the critical temperature
\begin{equation}
T_{\star}=\sqrt{\frac{12\bar{v}^{2}}{N+2}},
\end{equation}
where $m^{2}=-\lambda\bar{v}^{2}/6$ and the over-bar denotes the
zero temperature value of a quantity. That $T_{\star}$ is independent
of $\xi$ is consistent with the previously known result that the
same critical point is found in all symmetry improvement schemes \citep{Brown2015a}.
There is no subtlety involved in the $\mathrm{V}\beta\to\infty$ limit
in this case.

\subsection{Broken phase with $m_{G}^{2}\protect\neq0$\label{subsec:Broken-Phase-with-massive-Goldstones}}

Attempting to describe the broken phase with the SSI equations of
motion is rather more complicated than the symmetric phase. Decreasing
temperature at \emph{fixed} $\xi$ gives $\hat{\xi}\to0$ and
\begin{equation}
\mathcal{S}_{1}\to\left(\frac{4v^{2}}{\xi m_{G}^{2}\left(\mathrm{V}\beta\right)^{2}}\right)^{1/3},
\end{equation}
so the equations of motion \eqref{eq:SSI-2PI-HF-vev-eom}-\eqref{eq:SSI-2PI-HF-H-eom}
become for the vev

\begin{eqnarray}
0 & = & m^{2}+\frac{\lambda}{6}v^{2}\nonumber \\
 &  & +\left(N-1\right)\frac{\lambda}{6}\nonumber \\
 &  & \times\left[\frac{m_{G}^{2}}{16\pi^{2}}\ln\frac{m_{G}^{2}}{\mu^{2}}+\mathcal{T}_{G}^{\mathrm{th}}+\left(\frac{4v^{2}}{\xi m_{G}^{2}\left(\mathrm{V}\beta\right)^{2}}\right)^{1/3}\right]\nonumber \\
 &  & +\frac{\lambda}{2}\left(\frac{m_{H}^{2}}{16\pi^{2}}\ln\frac{m_{H}^{2}}{\mu^{2}}+\mathcal{T}_{H}^{\mathrm{th}}\right)\nonumber \\
 &  & +\frac{1}{\xi}\left(N-1\right)\left(\frac{\xi m_{G}^{2}}{\sqrt{2}v^{2}\mathrm{V}\beta}\right)^{2/3},
\end{eqnarray}

\begin{widetext}
and for the Goldstone

\begin{equation}
m_{G}^{2}=m^{2}+\frac{\lambda}{6}v^{2}+\left(N+1\right)\frac{\lambda}{6}\left[\frac{m_{G}^{2}}{16\pi^{2}}\ln\frac{m_{G}^{2}}{\mu^{2}}+\mathcal{T}_{G}^{\mathrm{th}}+\left(\frac{4v^{2}}{\xi m_{G}^{2}\left(\mathrm{V}\beta\right)^{2}}\right)^{1/3}\right]+\frac{\lambda}{6}\left(\frac{m_{H}^{2}}{16\pi^{2}}\ln\frac{m_{H}^{2}}{\mu^{2}}+\mathcal{T}_{H}^{\mathrm{th}}\right),
\end{equation}
and for the Higgs

\begin{equation}
m_{H}^{2}=m^{2}+\frac{\lambda}{2}v^{2}+\left(N-1\right)\frac{\lambda}{6}\left[\frac{m_{G}^{2}}{16\pi^{2}}\ln\frac{m_{G}^{2}}{\mu^{2}}+\mathcal{T}_{G}^{\mathrm{th}}+\left(\frac{4v^{2}}{\xi m_{G}^{2}\left(\mathrm{V}\beta\right)^{2}}\right)^{1/3}\right]+\frac{\lambda}{2}\left(\frac{m_{H}^{2}}{16\pi^{2}}\ln\frac{m_{H}^{2}}{\mu^{2}}+\mathcal{T}_{H}^{\mathrm{th}}\right).
\end{equation}
\end{widetext}

Note that all of the soft symmetry improvement terms vanish in the
limit $\mathrm{V}\beta\to\infty$. Thus the SSI-2PIEA reduces to the
unimproved 2PIEA if $\mathrm{V}\beta\to\infty$ at fixed $\xi$. It
is necessary to allow $\xi$ to vary as the $\mathrm{V}\beta\to\infty$
limit is taken to obtain a non-trivial correction to the unimproved
2PIEA. This is the first sign that the limit is non-trivial.

This section examines the simplest scheme to find a non-trivial limit.
This turns out to be the novel limit mentioned in the introduction.
It is shown in section \ref{subsec:Broken-Phase-with-massless-Goldstones}
that the only other non-trivial limit is equivalent to the old SI-2PIEA.
We proceed by letting $\xi$ vary with $\mathrm{V}\beta$ as $\xi=\left(\mathrm{V}\beta\right)^{\alpha}\zeta$
where $\zeta$ is a constant (with mass dimension $\left[\zeta\right]=2+4\alpha$).
If $\alpha\geq1$ the SSI terms vanish in the limit. If $\alpha<1$
\begin{equation}
\epsilon\to\left(\frac{\left(\mathrm{V}\beta\right)^{\alpha}\zeta}{4v^{2}\mathrm{V}\beta m_{G}^{4}}\right)^{1/3},
\end{equation}
and the symmetry improvement terms are

\begin{eqnarray}
\mathcal{S}_{1} & \to & \frac{1}{m_{G}^{2}}\left[\left(\frac{4v^{2}m_{G}^{4}}{\zeta}\right)^{1/3}\left(\mathrm{V}\beta\right)^{\left(-\alpha-2\right)/3}-\frac{1}{\mathrm{V}\beta}\right],\\
\mathcal{S}_{2} & \to & \frac{1}{\zeta}\left(N-1\right)\left(\frac{\zeta m_{G}^{2}}{\sqrt{2}v^{2}}\right)^{2/3}\left(\mathrm{V}\beta\right)^{\left(-2-\alpha\right)/3}.
\end{eqnarray}

The only non-trivial possibility is $\alpha=-2$, for which

\begin{eqnarray}
\epsilon & \to & \frac{1}{\mathrm{V}\beta}\left(\frac{\zeta}{4v^{2}m_{G}^{4}}\right)^{1/3}\to0,\\
\mathcal{S}_{1} & \to & \left(\frac{4v^{2}}{\zeta m_{G}^{2}}\right)^{1/3},\\
\mathcal{S}_{2} & \to & \frac{1}{\zeta}\left(N-1\right)\left(\frac{\zeta m_{G}^{2}}{\sqrt{2}v^{2}}\right)^{2/3}.
\end{eqnarray}
The equations of motion are for the vev

\begin{eqnarray}
0 & = & m^{2}+\frac{\lambda}{6}v^{2}\nonumber \\
 &  & +\left(N-1\right)\frac{\lambda}{6}\left[\frac{m_{G}^{2}}{16\pi^{2}}\ln\frac{m_{G}^{2}}{\mu^{2}}+\mathcal{T}_{G}^{\mathrm{th}}+\left(\frac{4v^{2}}{\zeta m_{G}^{2}}\right)^{1/3}\right]\nonumber \\
 &  & +\frac{\lambda}{2}\left(\frac{m_{H}^{2}}{16\pi^{2}}\ln\frac{m_{H}^{2}}{\mu^{2}}+\mathcal{T}_{H}^{\mathrm{th}}\right)\nonumber \\
 &  & +\left(N-1\right)\left(\frac{m_{G}^{2}}{\sqrt{2\zeta}v^{2}}\right)^{2/3},
\end{eqnarray}
for the Goldstone

\begin{eqnarray}
m_{G}^{2} & = & m^{2}+\frac{\lambda}{6}v^{2}\nonumber \\
 &  & +\left(N+1\right)\frac{\lambda}{6}\left[\frac{m_{G}^{2}}{16\pi^{2}}\ln\frac{m_{G}^{2}}{\mu^{2}}+\mathcal{T}_{G}^{\mathrm{th}}+\left(\frac{4v^{2}}{\zeta m_{G}^{2}}\right)^{1/3}\right]\nonumber \\
 &  & +\frac{\lambda}{6}\left(\frac{m_{H}^{2}}{16\pi^{2}}\ln\frac{m_{H}^{2}}{\mu^{2}}+\mathcal{T}_{H}^{\mathrm{th}}\right),
\end{eqnarray}
and for the Higgs

\begin{eqnarray}
m_{H}^{2} & = & m^{2}+\frac{\lambda}{2}v^{2}\nonumber \\
 &  & +\left(N-1\right)\frac{\lambda}{6}\left[\frac{m_{G}^{2}}{16\pi^{2}}\ln\frac{m_{G}^{2}}{\mu^{2}}+\mathcal{T}_{G}^{\mathrm{th}}+\left(\frac{4v^{2}}{\zeta m_{G}^{2}}\right)^{1/3}\right]\nonumber \\
 &  & +\frac{\lambda}{2}\left(\frac{m_{H}^{2}}{16\pi^{2}}\ln\frac{m_{H}^{2}}{\mu^{2}}+\mathcal{T}_{H}^{\mathrm{th}}\right).
\end{eqnarray}
Importantly, note that the mass appearing in Goldstone's theorem is
$\epsilon m_{G}^{2}$ (from the definition of $\epsilon$: $\Delta_{G}^{-1}\left(0,\boldsymbol{0}\right)=\epsilon m_{G}^{2}$),
which obeys $\epsilon m_{G}^{2}\to0$ as $\mathrm{V}\beta\to\infty$
thanks to the scaling chosen for $\xi$. Thus this scheme satisfies
Goldstone's theorem even if $m_{G}^{2}\neq0$. What $m_{G}^{2}\neq0$
indicates here is not actually a violation of Goldstone's theorem,
but a \emph{non}-communication of the masslessness of the Goldstone
zero mode to the other modes.

Defining the dimensionless variables

\begin{eqnarray}
x & = & v^{2}/\mu^{2},\\
y & = & m_{G}^{2}/\mu^{2},\\
z & = & m_{H}^{2}/\mu^{2},\\
\bar{x} & = & \bar{v}^{2}/\mu^{2},\\
\bar{X} & = & -6m^{2}/\lambda\mu^{2},\\
\hat{\zeta} & = & \zeta\mu^{6},\\
T_{G/H} & = & \mu^{-2}\mathcal{T}_{G/H}^{\mathrm{th}},
\end{eqnarray}
(note the distinction between the Lagrangian parameter $\bar{X}$
and the zero temperature value of the mean field $\bar{x}$, which
happen to be equal at tree level and in the usual renormalization
scheme for the Hartree-Fock approximation) this system becomes

\begin{eqnarray}
0 & = & \frac{\lambda}{6}\left(x-\bar{X}\right)\nonumber \\
 &  & +\left(N-1\right)\frac{\lambda}{6}\left[\frac{1}{16\pi^{2}}y\ln y+T_{G}+\left(\frac{4x}{\hat{\zeta}y}\right)^{1/3}\right]\nonumber \\
 &  & +\frac{\lambda}{2}\left(\frac{1}{16\pi^{2}}z\ln z+T_{H}\right)\nonumber \\
 &  & +\left(N-1\right)\left(\frac{y}{\sqrt{2\hat{\zeta}}x}\right)^{2/3},\label{eq:ssi-hf-veom-nondim}\\
y & = & \frac{\lambda}{6}\left(x-\bar{X}\right)\nonumber \\
 &  & +\left(N+1\right)\frac{\lambda}{6}\left[\frac{1}{16\pi^{2}}y\ln y+T_{G}+\left(\frac{4x}{\hat{\zeta}y}\right)^{1/3}\right]\nonumber \\
 &  & +\frac{\lambda}{6}\left(\frac{1}{16\pi^{2}}z\ln z+T_{H}\right),\label{eq:ssi-hf-geom-nondim}\\
z & = & \frac{\lambda}{3}x-\left(N-1\right)\left(\frac{y}{\sqrt{2\hat{\zeta}}x}\right)^{2/3}.\label{eq:ssi-hf-heom-nondim}
\end{eqnarray}

Looking for a zero temperature solution gives the system

\begin{eqnarray}
0 & = & \frac{\lambda}{6}\left(\bar{x}-\bar{X}\right)\nonumber \\
 &  & +\left(N-1\right)\frac{\lambda}{6}\left[\frac{1}{16\pi^{2}}\bar{y}\ln\bar{y}+\left(\frac{4\bar{x}}{\hat{\zeta}\bar{y}}\right)^{1/3}\right]\nonumber \\
 &  & +\frac{\lambda}{2}\frac{1}{16\pi^{2}}\bar{z}\ln\bar{z}+\left(N-1\right)\left(\frac{\bar{y}^{2}}{2\hat{\zeta}\bar{x}^{2}}\right)^{1/3},\label{eq:ssi-hf-veom-nondim-zerot}\\
\bar{y} & = & \frac{\lambda}{6}\left(\bar{x}-\bar{X}\right)\nonumber \\
 &  & +\left(N+1\right)\frac{\lambda}{6}\left[\frac{1}{16\pi^{2}}\bar{y}\ln\bar{y}+\left(\frac{4\bar{x}}{\hat{\zeta}\bar{y}}\right)^{1/3}\right]\nonumber \\
 &  & +\frac{\lambda}{6}\frac{1}{16\pi^{2}}\bar{z}\ln\bar{z},\label{eq:ssi-hf-geom-nondim-zerot}\\
\bar{z} & = & \frac{\lambda\bar{x}}{3}-\left(N-1\right)\left(\frac{\bar{y}^{2}}{2\hat{\zeta}\bar{x}^{2}}\right)^{1/3}.\label{eq:ssi-hf-heom-nondim-zerot}
\end{eqnarray}
First, ignoring the SSI terms, one finds the usual unimproved 2PI
solution $\bar{x}=\bar{X}$, $\bar{y}=0$, $\bar{z}=\lambda\bar{x}/3=1$.
Now examine the large $N$ limit of these equations, taking as the
scaling limit $g=\lambda N=\text{constant}$ and $\bar{x},\bar{X}\sim N^{1}$,
$\bar{y},\bar{z}\sim N^{0}$ and $\hat{\zeta}\sim N^{a}$ with $a$
to be determined. To leading order
\begin{eqnarray}
0 & = & \frac{g}{6N}\left(\bar{x}-\bar{X}\right)\nonumber \\
 &  & +\frac{g}{6}\left\{ \frac{1}{16\pi^{2}}\bar{y}\ln\bar{y}+N^{\left(1-a\right)/3}\left[\frac{4\bar{x}/N}{\left(\hat{\zeta}/N^{a}\right)\bar{y}}\right]^{1/3}\right\} \nonumber \\
 &  & +N^{\left(1-a\right)/3}\left[\frac{\bar{y}^{2}}{2\left(\hat{\zeta}/N^{a}\right)\left(\bar{x}/N\right)^{2}}\right]^{1/3},\\
\bar{y} & = & \frac{g}{6N}\left(\bar{x}-\bar{X}\right)+\left\{ \frac{g}{6}\frac{1}{16\pi^{2}}\bar{y}\ln\bar{y}\vphantom{\left[\frac{4\bar{x}/N}{\left(\hat{\zeta}/N^{a}\right)\bar{y}}\right]^{1/3}}\right.\nonumber \\
 &  & +\left.N^{\left(1-a\right)/3}\left[\frac{4\bar{x}/N}{\left(\hat{\zeta}/N^{a}\right)\bar{y}}\right]^{1/3}\right\} ,\\
\bar{z} & = & \frac{g\bar{x}}{3N}-N^{\left(1-a\right)/3}\left[\frac{\bar{y}^{2}}{2\left(\hat{\zeta}/N^{a}\right)\left(\bar{x}/N\right)^{2}}\right]^{1/3}.
\end{eqnarray}
Note that the $z$ dependence of the first two equations is higher
order in $1/N$. Scaling limits exist if $a\geq1$. Note that the
SSI term in $\Gamma_{\xi}^{\mathrm{SSI}}$ goes as $\xi^{-1}Nv^{2}\sim N^{3-a}$
so that one needs $a\geq2$ for a scaling limit for $\Gamma_{\xi}^{\mathrm{SSI}}$
to exist. $a=1$ can also be ruled out by considering the equations
of motion, for in this case the leading approximation is

\begin{eqnarray}
0 & = & \frac{g}{6N}\left(\bar{x}-\bar{X}\right)\nonumber \\
 &  & +\frac{g}{6}\left\{ \frac{1}{16\pi^{2}}\bar{y}\ln\bar{y}+\left[\frac{4\bar{x}/N}{\left(\hat{\zeta}/N\right)\bar{y}}\right]^{1/3}\right\} \nonumber \\
 &  & +\left[\frac{\bar{y}^{2}}{2\left(\hat{\zeta}/N\right)\left(\bar{x}/N\right)^{2}}\right]^{1/3},\label{eq:large-N-vev-eq}\\
\bar{y} & = & \frac{g}{6N}\left(\bar{x}-\bar{X}\right)\nonumber \\
 &  & +\frac{g}{6}\left\{ \frac{1}{16\pi^{2}}\bar{y}\ln\bar{y}+\left[\frac{4\bar{x}/N}{\left(\hat{\zeta}/N\right)\bar{y}}\right]^{1/3}\right\} ,\label{eq:large-N-m_G-eom}\\
\bar{z} & = & \frac{g\bar{x}}{3N}-\left[\frac{\bar{y}^{2}}{2\left(\hat{\zeta}/N\right)\left(\bar{x}/N\right)^{2}}\right]^{1/3}.\label{eq:large-N-m_H-eom}
\end{eqnarray}
Using \eqref{eq:large-N-vev-eq} to simplify \eqref{eq:large-N-m_G-eom},
\begin{equation}
\bar{y}=-\left(\frac{\bar{y}^{2}}{2\left(\hat{\zeta}/N\right)\left(\bar{x}/N\right)^{2}}\right)^{1/3},
\end{equation}
which has the solution

\begin{equation}
\bar{y}=-\frac{1}{2\left(\hat{\zeta}/N\right)\left(\bar{x}/N\right)^{2}}<0,
\end{equation}
with an unphysical tachyonic Goldstone $m_{G}^{2}<0$. This is not
\emph{necessarily} a problem because the zero mode $\Delta_{G}\left(0,\boldsymbol{0}\right)$
is always positive and, in finite volume with $\beta$ and $L$ sufficiently
small (i.e. $\beta,L<2\pi/\left|m_{G}\right|$), each mode $\Delta_{G}\left(n,\boldsymbol{k}\right)=1/\left(\omega_{n}^{2}+\boldsymbol{k}^{2}+m_{G}^{2}\right)$
with $n,\boldsymbol{k}\neq\boldsymbol{0}$ is still positive. Physically,
confinement energy is stabilizing the tachyon. However, a second condition
is that the imaginary part of the first equation of motion vanishes,
yielding
\begin{eqnarray}
0 & = & -\frac{1}{16\pi^{2}}\left|\bar{y}\right|\left(2k+1\right)\pi\nonumber \\
 &  & -\sin\left[\frac{\left(2k+1\right)\pi}{3}\right]\left[\frac{4\bar{x}/N}{\left(\hat{\zeta}/N\right)\left|\bar{y}\right|}\right]^{1/3},
\end{eqnarray}
where the branch chosen is $\bar{y}=\left|\bar{y}\right|\exp\left(i\pi\left(2k+1\right)\right)$
where $k$ is an integer. Using the solution for $\bar{y}$ this becomes

\begin{equation}
k=-\frac{1}{2}-32\pi\left(\hat{\zeta}/N\right)\left(\bar{x}/N\right)^{3}\sin\left(\frac{\left(2k+1\right)\pi}{3}\right),
\end{equation}
which only has solutions of the form $k=3j$ if
\begin{equation}
j=-\frac{1}{6}-\frac{16\pi}{\sqrt{3}}\left(\hat{\zeta}/N\right)\left(\bar{x}/N\right)^{3}
\end{equation}
is an integer. The existence of solutions only for certain discrete
values of $\hat{\zeta}\bar{x}^{3}$ is troubling and highly counter-intuitive
(note especially that the relationship between $\hat{\zeta}$ and
$\bar{x}$ for a given $j$ is independent of $g$, so that no matter
how $g$ is varied at fixed $m^{2}$ and $\hat{\zeta}$ , $\bar{x}$
is fixed even though one expects $\bar{x}\sim N/g$).

If $a>1$ the SSI terms in the equation of motion are of higher order
and the leading large $N$ approximation is just the standard one,
i.e.
\begin{align}
0 & =\frac{g}{6N}\left(\bar{x}-\bar{X}\right)+\frac{g}{6}\frac{1}{16\pi^{2}}\bar{y}\ln\bar{y},\\
\bar{y} & =\frac{g}{6N}\left(\bar{x}-\bar{X}\right)+\frac{g}{6}\frac{1}{16\pi^{2}}\bar{y}\ln\bar{y},\\
\bar{z} & =\frac{g\bar{x}}{3N},
\end{align}
which has the solution $\bar{x}=\bar{X}$, $\bar{y}=0$ and $\bar{z}=\lambda\bar{x}/3$
as expected. Now note that if $1<a<4$ the SSI terms go as a fractional
power of $N$ between $N^{0}$ and $N^{-1}$ which cannot balance
any of the terms coming from diagrams, which all go as integer powers
of $N^{-1}$. This implies that, if $a>1$, it must be of the form
$a=4+3k$ where $k=0,1,2,\cdots$. The SSI terms then scale as $N^{-\left(1+k\right)}$
in the equation of motion and $N^{-1-3k}$ in $\Gamma_{\xi}^{\mathrm{SSI}}$.
Thus the SSI equations of motion possess a satisfactory leading large
$N$ limit, but only if the scaling is such that the SSI terms are
of higher order. This is the first sign that the SSI terms are problematic.

Now consider the case where symmetry improvement is only weakly imposed,
i.e. the SSI terms are a small perturbation. Intuitively this can
be achieved by taking $\hat{\zeta}$ sufficiently large. It is thus
natural to solve the equations of motion \eqref{eq:ssi-hf-veom-nondim-zerot}-\eqref{eq:ssi-hf-heom-nondim-zerot}
perturbatively in powers of $\hat{\zeta}^{-1/3}$ as $\hat{\zeta}\to\infty$.
Writing $\bar{x}=\bar{x}_{0}+\hat{\zeta}^{-1/3}\bar{x}_{1}+\hat{\zeta}^{-2/3}\bar{x}_{2}+\cdots$
and so on, the leading equations of motion are just the unimproved
2PI ones

\begin{eqnarray}
0 & = & \frac{\lambda}{6}\left(\bar{x}_{0}-\bar{X}\right)+\left(N-1\right)\frac{\lambda}{6}\frac{1}{16\pi^{2}}\bar{y}_{0}\ln\bar{y}_{0}\nonumber \\
 &  & +\frac{\lambda}{2}\frac{1}{16\pi^{2}}\bar{z}_{0}\ln\bar{z}_{0},\\
\bar{y}_{0} & = & \frac{\lambda}{6}\left(\bar{x}_{0}-\bar{X}\right)+\left(N+1\right)\frac{\lambda}{6}\frac{1}{16\pi^{2}}\bar{y}_{0}\ln\bar{y}_{0}\nonumber \\
 &  & +\frac{\lambda}{6}\frac{1}{16\pi^{2}}\bar{z}_{0}\ln\bar{z}_{0},\\
\bar{z}_{0} & = & \frac{\lambda\bar{x}_{0}}{3}.
\end{eqnarray}
The first order perturbation obeys a system of equations which can
be arranged as the matrix equation
\begin{multline}
\left(\begin{array}{ccc}
\frac{\lambda}{6} & \frac{\left(N-1\right)\lambda}{96\pi^{2}}\left(1+\ln\bar{y}_{0}\right) & \frac{\lambda}{32\pi^{2}}\left(1+\ln\bar{z}_{0}\right)\\
\frac{\lambda}{6} & \frac{\left(N+1\right)\lambda}{96\pi^{2}}\left(1+\ln\bar{y}_{0}\right)-1 & \frac{\lambda}{96\pi^{2}}\left(1+\ln\bar{z}_{0}\right)\\
\frac{\lambda}{3} & 0 & -1
\end{array}\right)\left(\begin{array}{c}
\bar{x}_{1}\\
\bar{y}_{1}\\
\bar{z}_{1}
\end{array}\right)\\
=\left(N-1\right)\left(\frac{\bar{y}_{0}^{2}}{2\bar{x}_{0}^{2}}\right)^{1/3}\left(\begin{array}{c}
-\frac{\lambda\bar{x}_{0}}{3\bar{y}_{0}}-1\\
-\frac{N+1}{N-1}\frac{\lambda\bar{x}_{0}}{3\bar{y}_{0}}\\
1
\end{array}\right).
\end{multline}
Note that this equation is singular in the limit $\bar{y}_{0}\to0$.
The solution for $\bar{y}_{1}$ in this limit is

\begin{equation}
\bar{y}_{1}\to-\frac{32\pi^{2}}{\ln\bar{y}_{0}}\left(\frac{\bar{x}_{0}}{2\bar{y}_{0}}\right)^{1/3}\to\infty.
\end{equation}

There is no sense in which the SSI terms are a small perturbation,
no matter the value of $\hat{\zeta}$. This can also be seen from
a direct examination of the full equations of motion. In the limit
$\bar{y}\to0$ the $\left(\frac{4\bar{x}}{\hat{\zeta}\bar{y}}\right)^{1/3}$
terms always dominate for any finite value of $\hat{\zeta}$. The
result is that the SSI solution must always have $\bar{y}\neq0$,
even at zero temperature. For the same reason a perturbation analysis
near the critical temperature also fails and, in fact, real valued
solutions do not exist in a ($\hat{\zeta}$ dependent) range of temperatures
beneath the critical temperature. Further, $m_{G}^{2}$ appears to
increase as the SSI terms are more strongly imposed. Physically, the
unimproved 2PI equations of motion ``would like to have'' a non-zero
Goldstone mass. When the mass of the zero mode is forced to vanish
the SSI-2PIEA adjusts by \emph{increasing} the mass of the other modes.
This can be verified by examining numerical solutions.

\begin{figure}
\subfloat[\label{fig:ssi-hf-soln-x}]{\includegraphics[width=1\columnwidth]{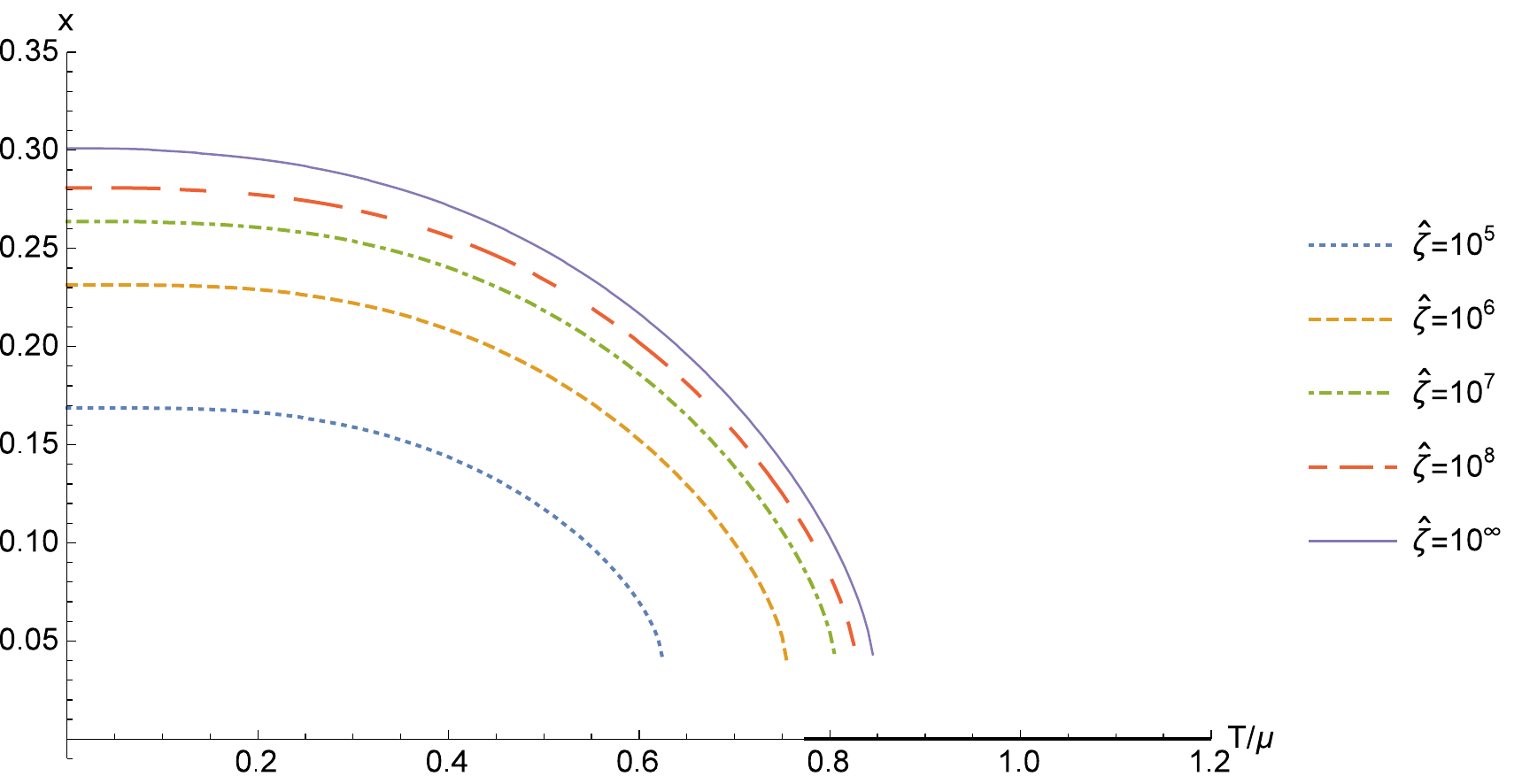}

}

\subfloat[\label{fig:ssi-hf-soln-y}]{\includegraphics[width=1\columnwidth]{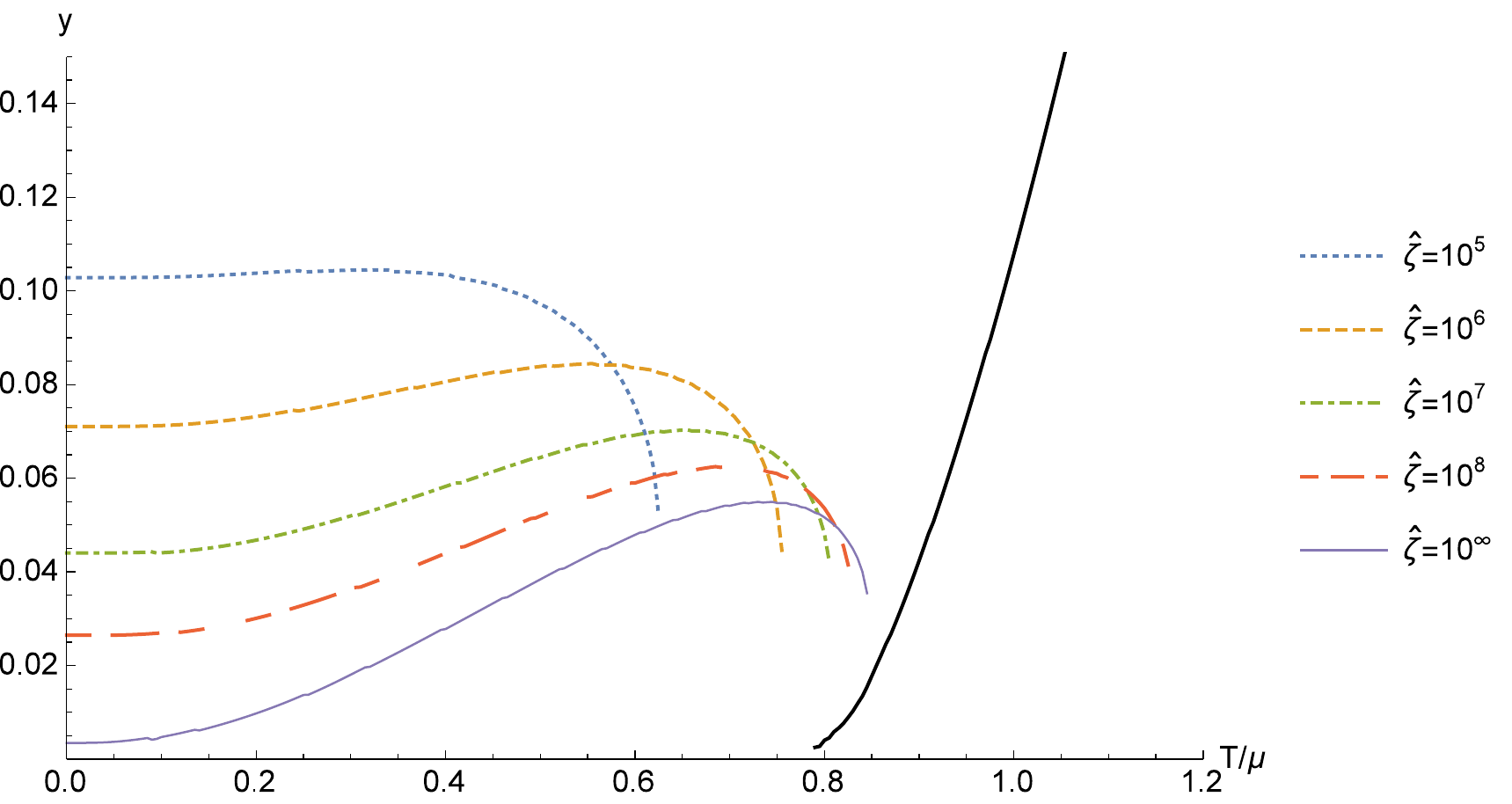}

}

\subfloat[\label{fig:ssi-hf-soln-z}]{\includegraphics[width=1\columnwidth]{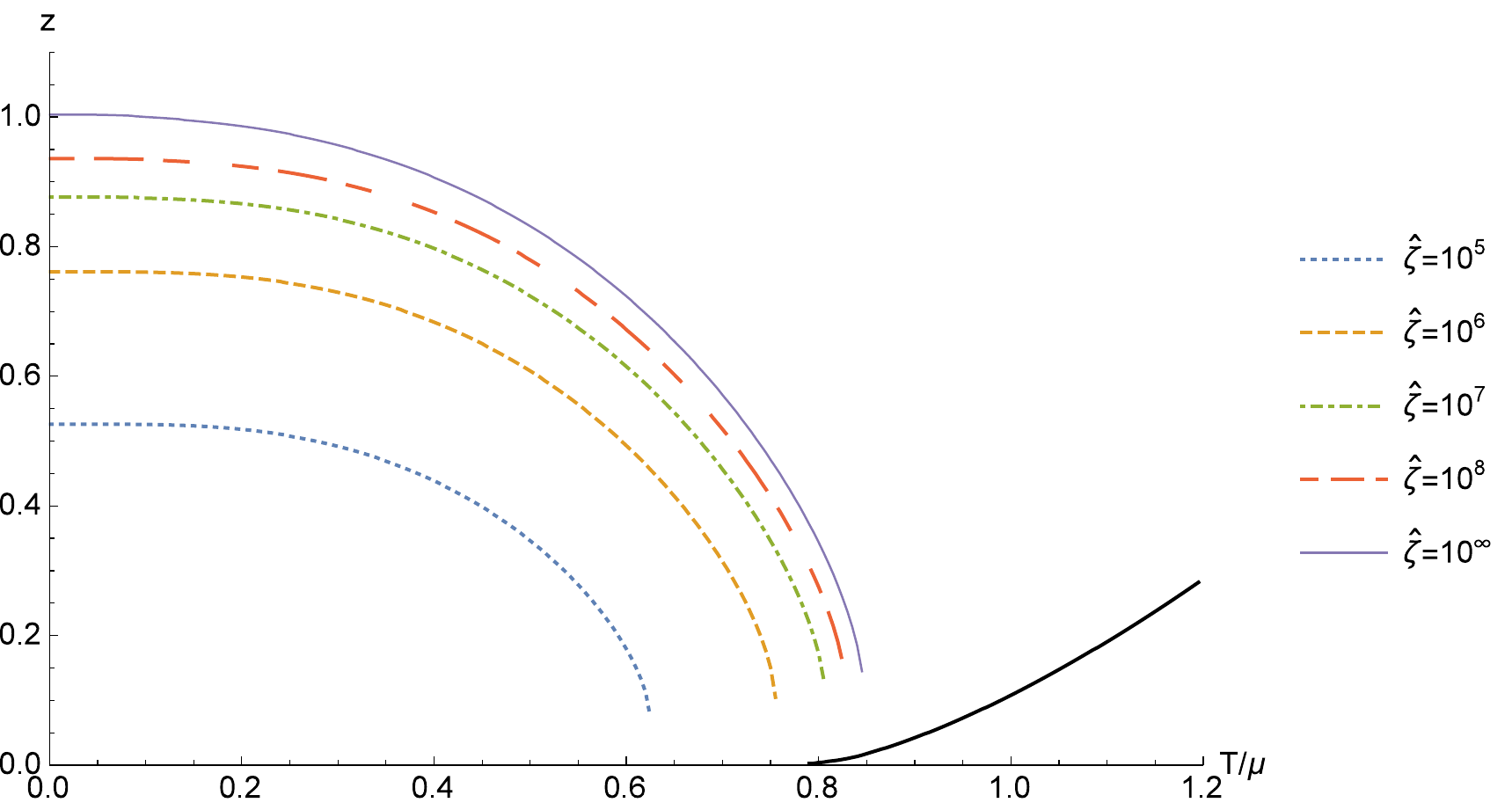}

}

\caption{\label{fig:ssi-hf-solns}(Color online) Solutions of \eqref{eq:ssi-hf-veom-nondim}-\eqref{eq:ssi-hf-heom-nondim},
the SSI equations of motion in the Hartree-Fock approximation. The
colored curves are broken phase $x$ (Fig \ref{fig:ssi-hf-soln-x}),
$y$ (Fig \ref{fig:ssi-hf-soln-y}) and $z$ (Fig \ref{fig:ssi-hf-soln-z})
and symmetric phase (solid black) solutions versus temperature for
$\lambda=10$, $N=4$, $\bar{X}=0.3$ and several values of $\hat{\zeta}$
from $10^{5}$ to $\infty$ (unimproved). The critical temperature
is $T_{\star}/\mu\approx0.775$.}
\end{figure}

\begin{figure}
\includegraphics[width=1\columnwidth]{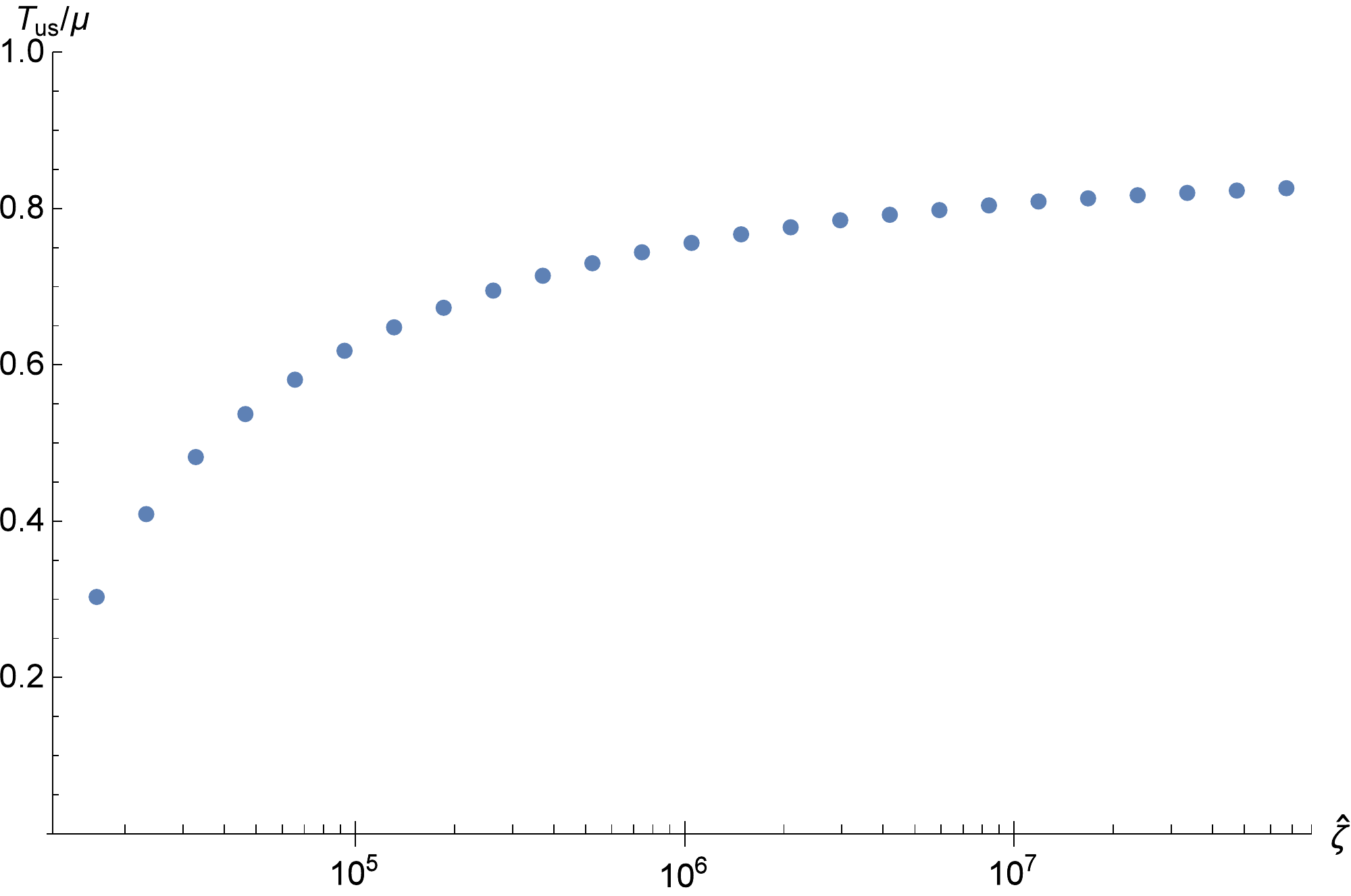}

\caption{\label{fig:ssi-hf-spinodal-temps}(Color online) The upper spinodal
temperature $T_{\mathrm{us}}$ versus $\hat{\zeta}$ for broken phase
solutions of the SSI equations of motion in the Hartree-Fock approximation
for $\lambda=10$, $N=4$, and $\bar{X}=0.3$. The critical temperature
is $T_{\star}/\mu\approx0.775$.}
\end{figure}

Numerical solutions of the \eqref{eq:ssi-hf-veom-nondim}-\eqref{eq:ssi-hf-heom-nondim}
are shown in Fig. \ref{fig:ssi-hf-solns} for $\lambda=10$, $N=4$,
$\bar{X}=0.3$ and several values of $\hat{\zeta}$ from $10^{4}$
to $\infty$. The critical temperature for these values is $T_{\star}/\mu\approx0.775$.
These parameter values are chosen for illustrative, not physical,
purposes. For very large $\hat{\zeta}$ the solution is near the unimproved
solution. However, as $\hat{\zeta}$ is decreased, $x$ and $z$ decrease
and $y$ \emph{increases} (this is consistent with the perturbation
$y_{1}$ being positive). Broken phase solutions cease to exist above
the upper spinodal temperature $T_{\mathrm{us}}\left(\hat{\zeta}\right)$
which depends on $\hat{\zeta}$. Note that $T_{\mathrm{us}}\left(\hat{\zeta}\right)$
drops below $T_{\star}$ for all $\hat{\zeta}<\hat{\zeta}_{c}$ where
$\hat{\zeta}_{c}$ is somewhere between $10^{6}$ and $10^{7}$. This
means that, for $\hat{\zeta}<\hat{\zeta}_{c}$ there is \emph{no solution}
between $T_{\mathrm{us}}\left(\hat{\zeta}\right)$ and $T_{\star}$.
Further, as $\hat{\zeta}\to0$, $T_{\mathrm{us}}\left(\hat{\zeta}\right)\to0$.
This behavior can be seen in Fig. \ref{fig:ssi-hf-spinodal-temps}.
At a critical value $\hat{\zeta}=\hat{\zeta}_{\star}$ one has $T_{\mathrm{us}}\left(\hat{\zeta}_{\star}\right)=0$
and real solutions cease to exist for all $\hat{\zeta}\leq\hat{\zeta}_{\star}$.

\begin{figure*}
\subfloat[]{\includegraphics[width=1\columnwidth]{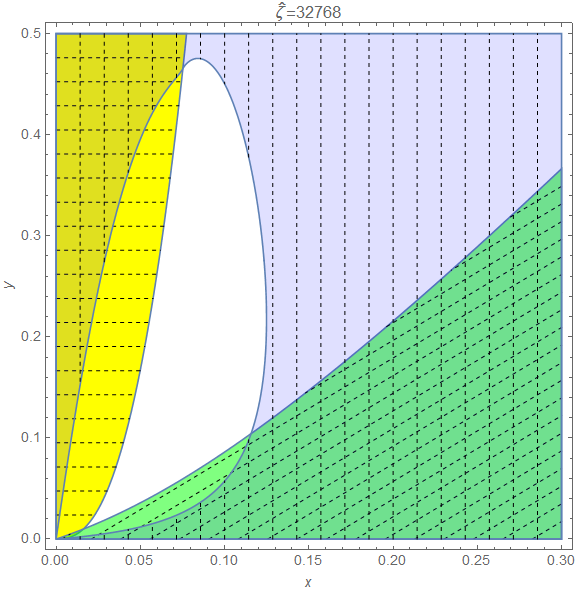}}\hfill{}\subfloat[]{\includegraphics[width=1\columnwidth]{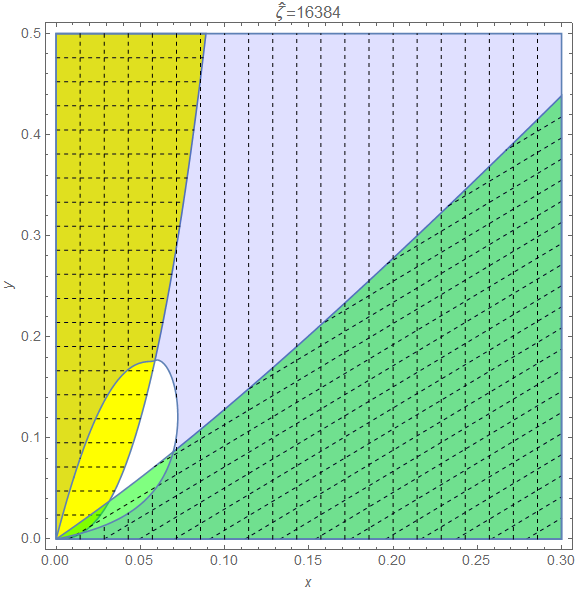}

}

\subfloat[]{\includegraphics[width=1\columnwidth]{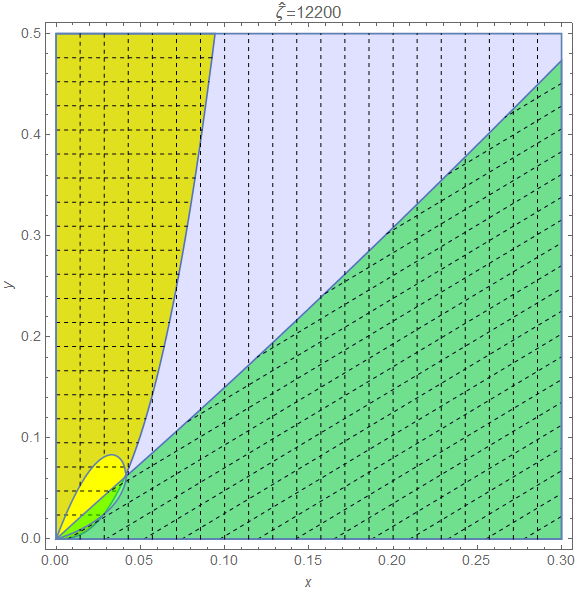}

}\hfill{}\subfloat[]{\includegraphics[width=1\columnwidth]{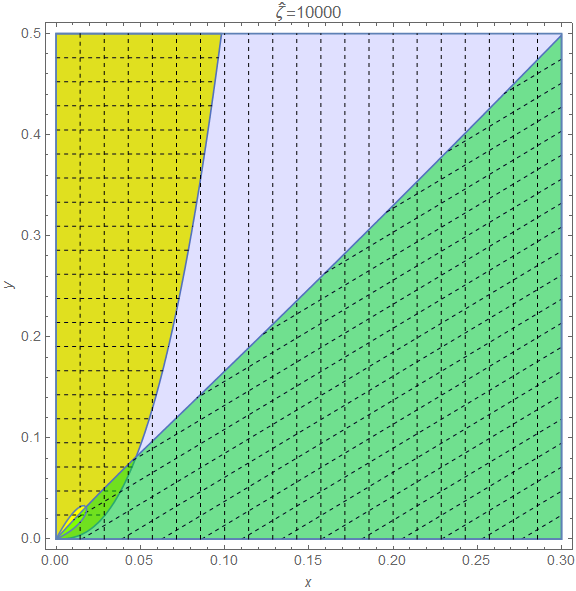}

}

\caption{\label{fig:loss-of-solution-plots}(Color online) Real and imaginary
parts of the right hand sides of \eqref{eq:ssi-hf-veom-nondim-zerot}
and \eqref{eq:ssi-hf-geom-nondim-zerot} as functions of $\bar{x}$
and $\bar{y}$ for $\lambda=10$, $N=4$, $\bar{X}=0.3$ and $\hat{\zeta}=2^{15}$
(upper left), $2^{14}$ (upper right), $12200$ (lower left) and $10^{4}$
(lower right). The blue vertically meshed regions satisfy $\Re\left(\eqref{eq:ssi-hf-veom-nondim-zerot}\right)>0$,
the yellow horizontally meshed regions satisfy $\Im\left(\eqref{eq:ssi-hf-veom-nondim-zerot}\right)\protect\neq0$
and the green diagonally meshed regions satisfy $\Re\left(\eqref{eq:ssi-hf-geom-nondim-zerot}\right)>\bar{y}$.}
\end{figure*}

The mathematical origin of this loss of solution can be understood
by considering the zero temperature equations of motion \eqref{eq:ssi-hf-veom-nondim-zerot}-\eqref{eq:ssi-hf-heom-nondim-zerot}.
We now use \eqref{eq:ssi-hf-heom-nondim-zerot} to eliminate $\bar{z}$
in the \eqref{eq:ssi-hf-veom-nondim-zerot} and \eqref{eq:ssi-hf-geom-nondim-zerot}
and consider the real and imaginary parts of the right hand sides
of these equations as functions of $\bar{x}$ and $\bar{y}$. The
relevant regions of the $\bar{x}-\bar{y}$ plane are shown in Fig.
\ref{fig:loss-of-solution-plots}. The blue vertically meshed regions
satisfy $\Re\left(\eqref{eq:ssi-hf-veom-nondim-zerot}\right)>0$,
yellow horizontally meshed regions satisfy $\Im\left(\eqref{eq:ssi-hf-veom-nondim-zerot}\right)\neq0$
and the green diagonally meshed regions satisfy $\Re\left(\eqref{eq:ssi-hf-geom-nondim-zerot}\right)>\bar{y}$.
Note that, for $\bar{x},\bar{y}>0$, $\Im\left(\eqref{eq:ssi-hf-veom-nondim-zerot}\right)\neq0$
is equivalent to $\bar{z}>0$, as is $\Im\left(\eqref{eq:ssi-hf-geom-nondim-zerot}\right)\neq0$
which does not give anything new.

Valid solutions of the equations of motion are on the boundary of
the blue and green regions simultaneously \emph{and} outside of the
yellow horizontally meshed region. As $\hat{\zeta}$ is decreased
it can be seen that the blue vertically meshed region ``closes in''
towards the origin, the green diagonally meshed region grows upwards,
and the yellow horizontally meshed region grows to the right. Solutions
cease to exist for $\hat{\zeta}=\hat{\zeta}_{\star}\approx12200$
where all three regions intersect at a common point. For all $\hat{\zeta}<\hat{\zeta}_{\star}$
there are no solutions (intersection points between the blue and green
curves) which are also real (outside the yellow horizontally meshed
region). If the temperature is non-zero, the thermal contributions
increase the real parts of the right hand sides which, in comparison
with Fig. \ref{fig:loss-of-solution-plots}, hastens the onset of
loss of solutions, which is therefore achieved at a greater value
$\hat{\zeta}$. Conversely, the temperature at which solutions are
lost for a given $\hat{\zeta}$ increases as a function of $\hat{\zeta}$,
which, of course, matches the behavior seen in Figs \ref{fig:ssi-hf-solns}
and \ref{fig:ssi-hf-spinodal-temps}.

\subsection{Broken phase with $m_{G}^{2}\to0$\label{subsec:Broken-Phase-with-massless-Goldstones}}

In order to find a broken phase solution without the pathological
properties of the previous section one can try to find solutions with
$m_{G}^{2}\to0$ in the $\mathrm{V}\beta\to\infty$ limit. To achieve
this, take the scalings

\begin{align}
\xi & =\left(\mathrm{V}\beta\right)^{\alpha}\zeta=\left(\mathrm{V}\beta\right)^{\alpha}\mu^{2+4\alpha}\hat{\zeta},\\
m_{G}^{2} & =\left(\mathrm{V}\beta\right)^{-\gamma}\mu^{2-4\gamma}y,
\end{align}
where $\gamma>0$. The definitions of the other dimensionless variables
($x$, $z$, etc.) are as before. Then

\begin{equation}
\epsilon\sim\begin{cases}
\left(\frac{\left(\mathrm{V}\beta\right)^{\alpha+2\gamma-1}}{\left(\mu^{-4}\right)^{\alpha+2\gamma-1}}\right)^{1/3}\left(\frac{\hat{\zeta}}{4xy^{2}}\right)^{1/3}, & \alpha+2\gamma-1<0,\\
1-\frac{\left(\mu^{-4}\right)^{\alpha+2\gamma-1}}{\left(\mathrm{V}\beta\right)^{\alpha+2\gamma-1}}\frac{4xy^{2}}{\hat{\zeta}}, & \alpha+2\gamma-1>0.
\end{cases}
\end{equation}
One can take the equations of motion \eqref{eq:SSI-2PI-HF-vev-eom}-\eqref{eq:SSI-2PI-HF-H-eom}
with the prescription $\mathcal{S}_{1}\to0$ because, as discussed
in section \ref{sec:Renormalisation-of-HF}, the Goldstone tadpole
reduces to the unmodified form in the massless case. The result is
the equation of motion for the vev

\begin{eqnarray}
0 & = & m^{2}+\frac{\lambda}{6}v^{2}+\left(N-1\right)\frac{\lambda}{6}\frac{T^{2}}{12}+\frac{\lambda}{2}\left(\frac{m_{H}^{2}}{16\pi^{2}}\ln\frac{m_{H}^{2}}{\mu^{2}}+\mathcal{T}_{H}^{\mathrm{th}}\right)\nonumber \\
 &  & +\mathcal{S}_{2},\label{eq:si-vev-eom}
\end{eqnarray}
for the Goldstone mass

\begin{equation}
0=m^{2}+\frac{\lambda}{6}v^{2}+\left(N+1\right)\frac{\lambda}{6}\frac{T^{2}}{12}+\frac{\lambda}{6}\left(\frac{m_{H}^{2}}{16\pi^{2}}\ln\frac{m_{H}^{2}}{\mu^{2}}+\mathcal{T}_{H}^{\mathrm{th}}\right),\label{eq:si-mg-eom}
\end{equation}
and for the Higgs mass

\begin{equation}
m_{H}^{2}=m^{2}+\frac{\lambda}{2}v^{2}+\left(N-1\right)\frac{\lambda}{6}\frac{T^{2}}{12}+\frac{\lambda}{2}\left(\frac{m_{H}^{2}}{16\pi^{2}}\ln\frac{m_{H}^{2}}{\mu^{2}}+\mathcal{T}_{H}^{\mathrm{th}}\right),\label{eq:si-mh-eom}
\end{equation}
having used that $\mathcal{T}_{G}^{\mathrm{fin}}=T^{2}/12$ for $m_{G}^{2}=0$.
Note that, remarkably, \eqref{eq:si-mg-eom}-\eqref{eq:si-mh-eom}
are nothing but the SI-2PIEA equations of motion (c.f. \citep{Brown2015a}).
The only thing new is the modification of the vev equation by the
term $\mathcal{S}_{2}$. To examine this further one must consider
the three cases $\alpha+2\gamma-1\gtreqless0$ which govern the possible
scaling behaviors of this term.

In the $\alpha+2\gamma-1>0$ case, $\epsilon\to1$ and
\begin{equation}
\mathcal{S}_{2}\to\mu^{2}\frac{\left(\mu^{-4}\right)^{\alpha+2\gamma}}{\left(\mathrm{V}\beta\right)^{\alpha+2\gamma}}\frac{\left(N-1\right)2y^{2}}{\hat{\zeta}}\to0.
\end{equation}
If, on the other hand, $\alpha+2\gamma-1=0$, $\epsilon$ is a constant
as $\mathrm{V}\beta\to\infty$ and
\begin{equation}
\mathcal{S}_{2}\to\mu^{2}\frac{\left(\mu^{-4}\right)}{\left(\mathrm{V}\beta\right)}\frac{1}{\hat{\zeta}}\left(N-1\right)2y^{2}\epsilon^{2}\to0.
\end{equation}
In both of these cases \eqref{eq:si-vev-eom} is unmodified by SSI
and cannot hold at the same time as the other two equations of motion.
To see this, solve the SI-2PI equations to get

\begin{equation}
m_{H}^{2}=-2m^{2}-\frac{1}{3}\lambda\left(N+2\right)\frac{T^{2}}{12},
\end{equation}
and
\begin{equation}
\frac{\lambda}{6}v^{2}=-m^{2}-\frac{1}{6}\left(N+1\right)\lambda\frac{T^{2}}{12}-\frac{1}{6}\lambda\left(\frac{m_{H}^{2}}{16\pi^{2}}\ln\frac{m_{H}^{2}}{\mu^{2}}+\mathcal{T}_{H}^{\mathrm{th}}\right).
\end{equation}
Now use these in \eqref{eq:si-vev-eom} to get

\begin{equation}
\frac{T^{2}}{12}=\frac{m_{H}^{2}}{16\pi^{2}}\ln\frac{m_{H}^{2}}{\mu^{2}}+\mathcal{T}_{H}^{\mathrm{th}},
\end{equation}
which only holds at $T=0$ (for $\mu=\bar{m}_{H}$) and $T=T_{\star}$.
There is no solution at any other temperature.

The remaining case is $\alpha+2\gamma-1<0$. For this case the SSI
term becomes
\begin{equation}
\mathcal{S}_{2}\to\mu^{2}\left[\frac{\left(\mu^{-4}\right)^{\alpha+2\gamma+2}}{\left(\mathrm{V}\beta\right)^{\alpha+2\gamma+2}}\right]^{1/3}\left(N-1\right)\left(\frac{y^{2}}{2\hat{\zeta}x^{2}}\right)^{1/3}.
\end{equation}

Looking for asymptotic balance, the only solution is $\alpha+2\gamma+2=0$
(which automatically satisfies the condition $\alpha+2\gamma-1<0$).
In this case \eqref{eq:si-vev-eom} reduces to (in terms of dimensionless
variables now)

\begin{eqnarray}
0 & = & -\frac{\lambda}{6}\bar{X}+\frac{\lambda}{6}x+\left(N-1\right)\frac{\lambda}{6}\frac{T^{2}/\mu^{2}}{12}+\frac{\lambda}{2}\left(\frac{z}{16\pi^{2}}\ln z+T_{H}\right)\nonumber \\
 &  & +\left(N-1\right)\left(\frac{y^{2}}{2\hat{\zeta}x^{2}}\right)^{1/3}.
\end{eqnarray}
Subtracting \eqref{eq:si-mh-eom} from this gives
\begin{equation}
\left(N-1\right)\left(\frac{y^{2}}{2\hat{\zeta}x^{2}}\right)^{1/3}=\frac{\lambda}{3}x-z,
\end{equation}
which can be easily solved for $y$, giving
\begin{equation}
y^{2}=2\hat{\zeta}x^{2}\left(\frac{\frac{\lambda}{3}x-z}{N-1}\right)^{3}.
\end{equation}
Note that $m_{G}^{2}=0$ regardless of the value of $y$. The only
constraint is that $0\leq y<\infty$ which requires $\lambda x/3\geq z$.
This can be verified using the solution of the SI-2PI equations of
motion
\begin{align}
z & =1-\frac{T^{2}}{T_{\star}^{2}},\\
x & =\bar{X}-\left[\left(N+1\right)\frac{T^{2}/\mu^{2}}{12}+T_{H}\right],
\end{align}
(recalling $\bar{z}=1$ and $\bar{X}=3/\lambda$ are the zero temperature
solutions and $T_{\star}^{2}=12\bar{X}\mu^{2}/\left(N+2\right)$)
so that

\begin{equation}
\frac{\lambda}{3}x-z=\frac{1}{\bar{X}}\left(\frac{T^{2}/\mu^{2}}{12}-T_{H}\right)\geq0,
\end{equation}
since the thermal integral $T_{H}$ is maximized for massless particles.
Thus $0\leq y^{2}<\infty$ and $y$ can always be chosen in $0\leq y<\infty$.
Thus all of the limits $\xi\sim\left(\mathrm{V}\beta\right)^{-2\gamma-2}$
and $m_{G}^{2}\sim\left(\mathrm{V}\beta\right)^{-\gamma}$ with $\gamma>0$
are equivalent and are identified as the unique limiting procedure
that gives back the old SI-2PIEA from the SSI-2PIEA.

One can also see that this procedure is the unique way of connecting
the SI and SSI methods by directly matching the SSI term in $\Gamma_{\xi}^{\mathrm{SSI}}$
with the Lagrange multiplier term in $\Gamma^{\mathrm{SI}}$. To do
this one must recall the original formulation of the symmetry improvement
method. The constraint term in the SI-2PIEA is (c.f. ``the simple
constraint'' discussed in \citep{Brown_2016})
\begin{equation}
\mathcal{C}=\frac{i}{2}\ell_{A}^{a}\mathcal{W}_{a}^{A}.
\end{equation}
The constraint is singular, meaning one must proceed by violating
the constraint by an amount $\sim\eta$ then taking a limit $\eta\to0$
such that $\ell\eta$ is a constant. In the previous literature \citep{Pilaftsis2013,Brown2015a,Brown_2016}
this procedure was carried out at the level of the equations of motion.
Now it is convenient to implement this at the level of the action
by shifting the constraint term to
\begin{equation}
\frac{i}{2}\ell_{A}^{a}\left(\mathcal{W}_{a}^{A}-i\mathcal{F}_{a}^{A}\right),
\end{equation}
where $\mathcal{F}_{a}^{A}\sim\eta$ is the regulator written in $\mathrm{O}\left(N\right)$-covariant
form. Setting the SI constraint term equal to the SSI term gives
\begin{equation}
\frac{i}{2}\ell_{A}^{a}\left(\mathcal{W}_{a}^{A}-i\mathcal{F}_{a}^{A}\right)=-\frac{1}{2\xi}\mathcal{W}_{a}^{A}\mathcal{W}_{a}^{A}.
\end{equation}

This can be simplified by recalling that $\mathcal{W}_{a}^{A}=\Delta_{ab}^{-1}T_{bc}^{A}\varphi_{c}$,
going to an anti-symmetric multi-index $A\to jk$ for the Lie algebra
indices, and using $T_{bc}^{jk}=i\left(\delta_{jb}\delta_{kc}-\delta_{jc}\delta_{kb}\right)$
and $\varphi_{c}=v\delta_{cN}$. This gives
\begin{multline}
2\frac{i}{2}\mathrm{V}\beta\ell_{cN}^{a}\left(iP_{ca}^{\perp}\left[\Delta_{G}^{-1}\left(0,\boldsymbol{0}\right)\right]v-i\mathcal{F}_{a}^{cN}\right)\\
=-\frac{1}{2\xi}\left(-2\left(N-1\right)v^{2}\mathrm{V}\beta\left[\Delta_{G}^{-1}\left(0,\boldsymbol{0}\right)\right]^{2}\right),
\end{multline}
having used $\int_{y}\Delta_{G}^{-1}\left(x,y\right)=\Delta_{G}^{-1}\left(0,\boldsymbol{0}\right)$
and introduced the transverse projector $P_{ca}^{\perp}=\delta_{ca}-\varphi_{c}\varphi_{a}/\varphi^{2}$.
Without loss of generality one can set
\begin{align}
\ell_{cN}^{a} & =P_{ac}^{\perp}\left(\frac{1}{N-1}\ell_{dN}^{d}\right),\\
\mathcal{F}_{a}^{cN} & =P_{ac}^{\perp}\mathcal{F},
\end{align}
and find
\begin{equation}
-\ell_{cN}^{c}\left(\Delta_{G}^{-1}\left(0,\boldsymbol{0}\right)v-\mathcal{F}\right)=\frac{1}{\xi}\left(N-1\right)v^{2}\left[\Delta_{G}^{-1}\left(0,\boldsymbol{0}\right)\right]^{2}.
\end{equation}
Now recall that the usual form of the SI regulator is $\Delta_{G}^{-1}\left(0,\boldsymbol{0}\right)v=m_{G}^{2}v=\eta m^{3}$
where $m$ is some arbitrary mass scale (it is convenient to take
$m=\mu$). This identifies $\mathcal{F}=\eta m^{3}$. The $\eta\to0$
limit is taken so that $\eta\ell_{cN}^{c}=\ell_{0}v$ is a constant.
Using this and $\Delta_{G}^{-1}\left(0,\boldsymbol{0}\right)=\epsilon m_{G}^{2}$
gives
\begin{equation}
-\frac{\ell_{0}v}{\eta}\left(\epsilon m_{G}^{2}v-\eta\mu^{3}\right)=\frac{1}{\xi}\left(N-1\right)v^{2}\left[\epsilon m_{G}^{2}\right]^{2}.
\end{equation}

It is now convenient to take $\eta=\left(\mathrm{V}\beta\right)^{-\delta}\mu^{-4\delta}$
with $\delta>0$. Taking also the usual scalings for $\xi$ and $m_{G}^{2}$,
one finds
\begin{equation}
-\left(\mathrm{V}\beta\right)^{\delta-\gamma}\mu^{4\left(\delta-\gamma\right)}\epsilon\ell_{0}xy+\ell_{0}\sqrt{x}=\frac{\mu^{-4\alpha-8\gamma}}{\left(\mathrm{V}\beta\right)^{\alpha+2\gamma}}\epsilon^{2}\left(N-1\right)\frac{xy^{2}}{\hat{\zeta}}.
\end{equation}
If $\alpha+2\gamma-1>0$ asymptotic balance is impossible (dominant
terms can be matched, but not subdominant terms). Likewise, balance
cannot be achieved for $\alpha+2\gamma-1=0$. If $\alpha+2\gamma-1<0$,
however,
\begin{widetext}
\begin{multline}
-\left(\mathrm{V}\beta\right)^{\delta-\gamma}\mu^{4\left(\delta-\gamma\right)}\left[\frac{\left(\mathrm{V}\beta\right)^{\alpha+2\gamma-1}}{\left(\mu^{-4}\right)^{\alpha+2\gamma-1}}\right]^{1/3}\left(\frac{\hat{\zeta}}{4xy^{2}}\right)^{1/3}\ell_{0}xy+\ell_{0}\sqrt{x}\\
=\frac{\mu^{-4\alpha-8\gamma}}{\left(\mathrm{V}\beta\right)^{\alpha+2\gamma}}\left[\frac{\left(\mathrm{V}\beta\right)^{\alpha+2\gamma-1}}{\left(\mu^{-4}\right)^{\alpha+2\gamma-1}}\right]^{2/3}\left(\frac{\hat{\zeta}}{4xy^{2}}\right)^{2/3}\left(N-1\right)\frac{xy^{2}}{\hat{\zeta}}.\label{eq:si-equiv-ssi-eq}
\end{multline}
\end{widetext}

Matching powers of $\left(\mathrm{V}\beta\right)$ on both sides gives
\begin{align}
0 & =3\delta-1+\alpha-\gamma,\\
0 & =\alpha+2\gamma+2,
\end{align}
which of course duplicates the previous result. These equations have
the solutions
\begin{align}
\alpha & =-2\delta,\\
\gamma & =\delta-1.
\end{align}
$\gamma>0$ requires $\delta>1$. Substituting this into \eqref{eq:si-equiv-ssi-eq}
gives

\begin{equation}
-\left(\frac{\hat{\zeta}}{4xy^{2}}\right)^{1/3}\ell_{0}xy+\ell_{0}\sqrt{x}=\left(\frac{\hat{\zeta}}{4xy^{2}}\right)^{2/3}\left(N-1\right)\frac{xy^{2}}{\hat{\zeta}},
\end{equation}
which can be solved for $\hat{\zeta}$, giving
\begin{equation}
\hat{\zeta}^{1/3}=\left(\frac{1}{2\sqrt{x}y}\right)^{1/3}\left(1\pm\sqrt{1-\left(N-1\right)\frac{y}{\ell_{0}}}\right).
\end{equation}
This is the desired connection between the SSI stiffness parameter
$\hat{\zeta}$ and the SI Lagrange multiplier $\ell_{0}$.

\section{\label{sec:Discussion}Discussion}

In this paper we have introduced a new method of \emph{soft symmetry
improvement} (SSI) which relaxes the constraint of the symmetry improvement
(SI) method. Violations of Ward identities (WIs) are allowed but punished
in the solution of the SSI effective action. The method is essentially
a least-squares implementation of the symmetry improvement idea. A
new parameter, the stiffness $\xi$, controls the strength of the
constraint. We studied the SSI-2PIEA for a scalar $\mathrm{O}\left(N\right)$
model in the Hartree-Fock approximation and found that the method
is IR sensitive. The system must be formulated in finite volume $\mathrm{V}$
and temperature $T=\beta^{-1}$ and the $\mathrm{V}\beta\to\infty$
limit taken carefully.

We found three distinct limits in section \ref{sec:Solution-in-the-infinite-volume-limit}.
In all cases the symmetric phase is the same and is unmodified from
either the unimproved 2PIEA or SI-2PIEA methods. Only the broken phase
is affected by SSI. Two of the limits are equivalent to the unimproved
2PIEA and SI-2PIEA respectively. The third is a new limit where $\hat{\zeta}=\left(\mathrm{V}\beta\right)^{2}\xi\mu^{6}$
is taken to be fixed and finite as $\mathrm{V}\beta\to\infty$. In
this limit the WI is satisfied but the phase transition is strongly
first order and strongly dependent on the scaled stiffness $\hat{\zeta}$.
Also, the upper spinodal temperature decreases as $\hat{\zeta}$ decreases
and, for $\hat{\zeta}<\hat{\zeta}_{c}$, solutions fail to exist between
the upper spinodal temperature and the critical temperature. For $\hat{\zeta}=\hat{\zeta}_{\star}$,
the upper spinodal temperature is equal to zero and broken phase solutions
cease to exist entirely. The limit was studied in both the leading
large $N$ limit and in perturbation theory in $\hat{\zeta}^{-1/3}$.
The large $N$ limit is trivial to leading order and the perturbation
theory does not exist since the SSI term is singular at the unimproved
solution. These results all suggest that the new limit is pathological.

The results of this paper are primarily restricted by the use of the
Hartree-Fock approximation. Investigations of higher order approximations
are motivated but would be far more involved, numerically, than anything
attempted here. It is possible that a higher order truncation could
ameliorate some or all of the problems with SSI found here. However,
assuming the Hartree-Fock results hold true, we can summarize the
findings as follows: We have found a method which subsumes both the
unimproved 2PIEA and SI-2PIEA and contains a new dynamical limit as
$\mathrm{V}\beta\to\infty$. However, these limits are disconnected
from each other; there is no smooth way to interpolate from one to
another. Further, each limit is in one way or other pathological.
These results suggest that any potential advantages of SSI methods
(and likely any consideration of (S)SI out of equilibrium) \emph{must
occur in finite volume}. Whether this is possible or not depends on
the particular system being studied. Thus, ultimately, symmetry improvement
methods cannot be trusted as a ``black box'': their validity must
be decided on a case by case basis.
\begin{acknowledgments}
We would like to thank Daniel S. Kosov and Ron D. White for helpful
comments.
\end{acknowledgments}

\appendix

\end{document}